\documentclass[reqno,11pt]{article}
\usepackage{amssymb,amsthm,amsmath,amsfonts}
\usepackage{epsfig}

\theoremstyle{plain}
\begingroup
 
\newtheorem{thm}{Theorem}[section] 
\newtheorem{cor}[thm]{Corollary}
\newtheorem{lemma}[thm]{Lemma}

\endgroup

\newcommand{\nwc}{\newcommand}
\nwc{\bit}{\begin{itemize}}
\nwc{\eit}{\end{itemize}}
\nwc{\Levy}{L\'evy}
\nwc{\LK}{L\'evy-Khintchine}
\nwc{\LI}{L\'evy-It\^{o}}
\nwc{\be}{\begin{equation}}
\nwc{\ee}{\end{equation}}
\nwc{\ba}{\begin{eqnarray}}
\nwc{\ea}{\end{eqnarray}}
\nwc{\la}{\label}
\nwc{\nn}{\nonumber}
\nwc{\Z}{\mathbb{Z}}
\nwc{\C}{\mathbb{C}}
\nwc{\E}{\mathbb{E}}
\nwc{\R}{\mathbb{R}}
\nwc{\N}{\mathbb{N}}
\nwc{\attr}{\mathcal{A}_{\rm p}}
\nwc{\attrx}{\mathcal{A}}
\nwc{\PP}{\mathcal{P}}
\nwc{\PPE}{\mathcal{P}(E)}
\nwc{\M}{\mathcal{M}}
\nwc{\Tt}{T^{(t)}}
\nwc{\Ut}{U^{(t)}}
\nwc{\gtx}{g^{(t)}_x}
\nwc{\law}{\stackrel{\mathcal{L}}{\rightarrow}}
\nwc{\eqd}{\stackrel{d}{=}}
\nwc{\vp}{\varphi}
\nwc{\Vp}{\Phi}
\nwc{\psilevy}{\Psi}
\nwc{\ve}{\varepsilon}
\nwc{\veps}{\varepsilon}
\nwc{\eps}{\ve}
\nwc{\betarsc}{\beta}
\nwc{\cl}{c\'{a}dl\'{a}g}
\nwc{\qref}[1]{(\ref{#1})}
\nwc{\D}{\partial}
\nwc{\Ebar}{{\bar{E}}}
\nwc{\ebar}{[0,\infty)}
\nwc{\mmt}{m}
\nwc{\fzero}{F_\rho} %{F_{0,\rho}}
\nwc{\fone}{M_\rho} %{F_{1,\rho}}
\nwc{\ip}[1]{\langle #1 \rangle}
\nwc{\ipbig}[1]{\left\langle #1 \right\rangle}
\nwc{\Lip}{\mathop{\rm Lip}\nolimits}
\nwc{\Tmin}{T_{\min}}
\nwc{\Tmax}{T_{\max}}
\nwc{\Tgel}{T_{\rm gel}}
\nwc{\LL}{\mathcal{L}}
\nwc{\mudot}{\mu}
\nwc{\nudot}{\nu}
\nwc{\rme}{{\rm e}}
\nwc{\rmi}{{\rm i}}
\nwc{\nsup}{^{(n)}}
\nwc{\nsupj}{^{(n_j)}}
\nwc{\ksup}{^{(k)}}
\nwc{\jsup}{^{(j)}}
\nwc{\tsup}{^{(t)}}
\nwc{\nksup}{^{(n_k)}}
\nwc{\inv}{^{-1}}
\nwc{\qxfac}{(1-\rme^{-qx})}
\nwc{\Mcan}{\mathcal{S}_d}
\nwc{\Mcanx}{\mathcal{S}}
\nwc{\sm}{G}
\nwc{\dsm}{H}
\nwc{\Hmap}{{\mathfrak S}_p}
\nwc{\Hmapx}{{\mathfrak S}}
\nwc{\canmap}{f}
\nwc{\canonical}{\mathcal{M}}
\nwc{\dnto}{\downarrow}
\nwc{\upto}{\uparrow}
\nwc{\cpair}{c}  % For Levy pair
\nwc{\intR}{\int_0^\infty}
\nwc{\tnu}{\tilde\nu}
\nwc{\smeas}{s-measure} %{$s$-measure}
\nwc{\smeass}{s-measures} %{$s$-measures}
\nwc{\barsmeas}{$\overline{\mbox{s}}$-measure} %{$\bar s$-measure}
\nwc{\barsmeass}{$\overline{\mbox{s}}$-measures} %{$\bar s$-measures}
\nwc{\nubar}{\bar{\nu}}
\nwc{\mubar}{\bar{\mu}}
\nwc{\dust}{a_0}
\nwc{\bdust}{b_0}
\nwc{\gel}{a_\infty}
\nwc{\bgel}{b_\infty}
\nwc{\ggel}{g_\infty}
\nwc{\Fbar}{\bar{F}}
\nwc{\Fcheck}{\check{F}}
\nwc{\nucheck}{\tilde{\nu}}
\nwc{\hvp}{\hat{\vp}}
\nwc{\hpsi}{\hat{\psi}}
\nwc{\htt}{\hat{t}}
\nwc{\hq}{\hat{q}}
\nwc{\hs}{\hat{s}}
\nwc{\dist}{\mathop{\rm dist}\nolimits}
\nwc{\extsol}{\Fbar}
\nwc{\specialF}{\hat\Fbar}
\nwc{\esm}{\bar{\sm}}
\nwc{\tlam}{\tilde\lambda}
\nwc{\lgr}{a} 
\nwc{\Fgr}{F_{\rho,\gamma}}
\nwc{\vpgr}{\vp_{\rho,\gamma}}
\renewcommand{\Re}{\mathop{\rm Re}\nolimits}

\theoremstyle{definition}
\newtheorem{defn}{Definition}[section]
\newtheorem{rem}[defn]{Remark}

\theoremstyle{remark}

\numberwithin{equation}{section}
\numberwithin{figure}{section}
\begin{document}
\title{The scaling attractor and ultimate dynamics for Smoluchowski's
  coagulation equations} 
\author{Govind Menon\textsuperscript{1} and Robert. L.
Pego\textsuperscript{2}}

\date{\today}

\maketitle

\begin{abstract}
We describe a basic framework for studying dynamic scaling
that has roots in dynamical systems and probability theory.
Within this framework, we study Smoluchowski's coagulation equation 
for the three simplest rate kernels $K(x,y)=2$, $x+y$ and $xy$. 
In another work,
we classified all self-similar solutions and all universality
classes (domains of attraction) for scaling limits under weak
convergence (Comm.\ Pure Appl.\ Math 57 (2004)1197-1232).  Here
we add to this a complete description of the set of all limit
points of solutions modulo scaling (the {\it  scaling attractor})
and the dynamics on this limit set (the {\em ultimate
dynamics\/}). The main tool is Bertoin's \LK\/ representation
formula for eternal solutions of Smoluchowski's equation (Adv.\
Appl.\ Prob.\ 12 (2002) 547--64).  This representation 
{\em linearizes} the dynamics on the scaling attractor, 
revealing these dynamics to be
conjugate to a continuous dilation, and chaotic in a classical sense.
Furthermore, our study of scaling limits explains how Smoluchowski dynamics 
``compactifies'' in a natural way that accounts for clusters 
of zero and infinite size (dust and gel). 
\end{abstract} 



\medskip

\footnotetext[1]
{Division of Applied Mathematics, Box F, Brown University, Providence, RI 02912.
Email: menon@dam.brown.edu}
\footnotetext[2]{Department of Mathematical Sciences, 
Carnegie Mellon University, Pittsburgh, PA 15213.
Email: rpego@cmu.edu}

\pagebreak
\section{Introduction}
\label{sec:intro}
Smoluchowski's coagulation equation is a fundamental mean-field model
of clustering processes. The merging of clusters of mass $x$ and mass
$y$ to produce clusters of mass $x+y$ 
occurs at a mass-action rate modulated by a symmetric rate kernel
$K(x,y)$. Formally, the evolution equation for the density $n(t,x)$
of the size distribution reads
\begin{eqnarray}
{\partial_t n}(t,x) &=& 
\frac{1}{2} \int_0^x K(x-y,y) n(t,x-y) n(t,y) dy  
\nonumber\\
&&\ -\ \int_0^\infty K(x,y) n(t,x) n(t,y) dy,
\label{eq:smol1}
\end{eqnarray}
Many kernels arising in applications are homogeneous, that is, there
is $\gamma$ such that  $K(\alpha x, \alpha y) = \alpha^\gamma K(x,y)$
for every $\alpha,x,y >0$.  We restrict attention to the ``solvable''
kernels $K(x,y)=2$, $x+y$ and $xy$ ($\gamma=0$, 1 and 2 respectively)
which may be
studied via the Laplace transform. These special kernels arise in a
variety of applications, 
including the aggregation of colloids \cite{Chandra}($K=2$), 
droplet formation in clouds \cite{Golovin},
%gravitational clustering \cite{},
a phase transition for parking \cite{Chassaing},
Burgers' model of turbulence \cite{B_burgers} ($K=x+y$),
random graphs~\cite{Aldous} ($K=xy$), and kinetics of
polymerization~\cite{Ziff} (all kernels).

As time proceeds, the typical cluster size grows, and
an issue of relevance for homogeneous kernels is whether and how the size  
distribution develops toward self-similar form. In the
physics literature, this is called the {\em dynamic scaling\/} problem.  
This work continues our study of dynamic scaling for Smoluchowski's 
equation for the solvable kernels. We lay out a
general framework for the analysis of dynamic scaling that is 
inspired by elements from both dynamical systems and probability
theory. The main issues may be set out as a list of basic and
general questions:
\bit
\item What scaling solutions exist? 
Here we seek {\em self-similar solutions\/}, 
or fixed points of the dynamics modulo scaling.
\item What are the domains of attraction of these scaling solutions? 
These comprise the {\em universality classes\/} for dynamic scaling.
\item What limit points are possible under scaling dynamics in general?
We call the set of such points the {\it scaling attractor} of the system.
\item How can we describe the dynamics on the scaling attractor? 
We call this the {\it ultimate dynamics} of the system.
\item How complicated can the ultimate dynamics be?
\eit

While evidently stated in dynamical terms, strong motivation for this
framework comes from the classical limit theorems of probability theory
developed by the pioneers of the subject in the 1930s. 
These theorems concern limits of scaled sums of independent and
identically distributed random variables $X_1$, $X_2$, $\ldots$ with
common distribution $F$. 
The distribution of $Y_n=\sum_{j=1}^n X_j$ is the $n$-fold convolution
of $F$ with itself and this can be regarded as a discrete evolution
law for $F$.  The theory as laid out in the marvelous book of Feller
\cite{Feller} provides complete answers to the questions above:
\bit
\item {\em Scaling solutions.}
The normal distribution is the unique invariant distribution of finite
variance.  More generally, all 
invariant distributions (including those with heavy tails) 
comprise a two-parameter family, the {\em stable laws\/}
first  characterized completely by L\'{e}vy (see~\cite{Loeve} for a
historical account). 
%The densities for these distributions may be
%written as $p(x;\alpha,\theta)$, where $x \in \R$, $0 < \alpha \leq
%2$, $|\theta| \leq \min\{\alpha, 2-\alpha\}$. The parameter $\alpha$
%measures the algebraic decay of $p(x;\alpha,\theta)$ as $|x| \to
%\infty$. At the endpoint $\alpha=2$ we recover the normal distribution.   
\item {\em Domains of attraction.}
The central limit theorem states that the normal law 
attracts all distributions with finite variance. The domains of
attraction of the stable laws are completely 
classified in terms of the  power-law
behavior (more precisely, {\em regular variation\/}) and skewness of
the 2nd-moment distribution function. There 
are no other invariant distributions or domains of attraction in
the limit $n\to\infty$.
\item {\em Infinite divisibility.}
The most general distributions that can arise 
as rescaled limits for some subsequence $n_j\to\infty$ are the 
{\em infinitely divisible distributions\/}, characterized
by the famous \LK\/ formula.
The characteristic function $\varphi(k)
= \mathbb{E}(e^{ikX})$ of an infinitely divisible random variable $X$
is of the form $\varphi(k) = \exp\Psi(k)$ where 
\[\Psi(k) = 
\int_\mathbb{R} \frac{(e^{ikx}-
1 -ik\sin x)}{x^2} M(dx) + ibk 
\]  
with $\int_{\R} (1\wedge x^{-2}) M(dx) < \infty$ and $b\in\R$.
The measure $M$ is called the {\em canonical measure}.
\item {\em Ultimate dynamics.}
Under appropriate rescaling, nondegenerate limit points for the 
discrete evolution $n\mapsto S_n$ correspond one-to-one to 
 {\it \Levy\/ processes}  (continuous-time random walks with stationary
  independent increments).  
These are stochastic processes $X_t$ ($t>0$)
with right continuous paths with left limits,
 obeying the semigroup formula 
$\varphi_t(k) = \mathbb{E}(e^{ikX_t}) = e^{t\Psi(k)}$. 
Rescaling amplitude and $t$ (scaling dynamics) corresponds to 
linear tranformations (dilation, stretching, shifts) of the canonical measure.
%[Stop here? If not, strictly
%  speaking, one should describe the full \LK\/ representation
%  including the Gaussian coefficient and mean and not just the jump
%  measure]  in terms of the \Levy\
%measure, which is the jump-size distribution for a continuous-time random walk
%whose time-one increments follow the given infinitely divisible
%law. 
\item {\em Chaos.}
There exist distributions ({\it Doeblin's universal laws})
for which {\it every} infinitely divisible distribution appears as
some limit point for rescaled self-convolutions.  
%Thus the ultimate dynamics among the infinitely divisible laws is chaotic
%in a sense.
\eit

In an earlier article \cite{MP1}, we characterized the approach to
self-similarity for Smoluchowski's equation as $t\to\infty$ for $K=2$
and $x+y$, and the approach to self-similar blow-up as 
$t$ approaches the {\em gelation time} for $K=xy$.
By way of addressing the first two issues in the framework above we found:  
\bit
\item 
For each solvable kernel, with degree of homogeneity $\gamma\in\{0,1,2\}$, there
is a (classically known) self-similar solution with finite
$\gamma+1$st 
moment, unique up to normalization. This solution has
exponential decay as $x \to \infty$.
But more generally, there is a 
one-parameter family of self-similar solutions, corresponding to 
distribution functions 
written $F_{\rho,\gamma}$ for $\rho \in (0,1], \gamma \in \{0,1,2\}$. 
We recover the classical self-similar solution at the endpoint $\rho=1$. 
For $0 < \rho < 1$, the $F_{\rho,\gamma}$ have 
infinite $\gamma+1$st moment and are directly related 
to important heavy-tailed distributions of probability theory --- 
Mittag-Leffler distributions for $K=2$, and stable laws of maximum skewness 
for $K=x+y$ and $xy$. For $K=x+y$ these solutions were first discovered by
Bertoin by a different argument~\cite{B_eternal}.

\item The classical self-similar solution with $\rho=1$ attracts all
solutions with finite $\gamma+1$st moment. 
In general, the domains of attraction of self-similar solutions 
are characterized by
the regular variation of the $\gamma+1$st moment distribution. A positive
measure $\nu_0$ lies in the domain of attraction of the self-similar
solution $F_{\rho,\gamma}$ if and only if
\begin{equation}\label{1.domain}
\int_0^x y^{\gamma+1}\nu_0(dy) \sim x^{1-\rho}L(x), \qquad x\to \infty,
\end{equation}
for some function $L$ slowly varying at $\infty$ ($L(\lambda x)/L(\lambda)\to1$
as $\lambda\to\infty$ for all $x>0$). There are no other self-similar
solutions or domains of attraction.
\eit

Our present goal is to investigate more generally the 
scaling dynamics of Smoluchowski's equation for the solvable
kernels, as $t\to\infty$ for $K=2$ and $x+y$ and as $t$ approaches
the gelation time for $K=xy$. 
To stress the probabilistic analogy, if our earlier
article was a study of stable laws, here we study infinite
divisibility. The basis for our work is Bertoin's characterization 
of {\it  eternal solutions\/} 
for Smoluchowski's equation with additive kernel
$K=x+y$~\cite{B_eternal}. Eternal solutions for this kernel are defined 
for all $t\in(-\infty,\infty)$
(i.e., they may be extended  {\em backwards\/} in time globally), 
and thus they are `infinitely divisible' under clustering dynamics.
Bertoin established a remarkable  \LK-type representation for these
solutions. We generalize this result to other solvable kernels, and
revisit it from a dynamical systems viewpoint. In the context of the
framework above, we find that
\bit
\item The (proper) scaling attractor corresponds in one-to-one fashion with
eternal solutions of Smoluchowski's equation, and these have a
\LK\/ representation for each solvable kernel.
\item The \LK\/ representation {\em linearizes the dynamics on the
  attractor}.  Nonlinear evolution by Smoluchowski's equation on the attractor  
  is conjugate to a group of simple scaling transformations on the 
  measures that generate the representation.
 Heuristically, the classification of the domains of
attraction in~\cite{MP1} shows 
that the dynamics are extremely sensitive to the tails of
the initial size distribution. The  \LK\/ representation
makes this precise: the ultimate dynamics on the
  scaling attractor is conjugate to a continuous dilation map.
\item One may use the \LK\ representation to construct orbits with
  complicated dynamics. The scaling attractor contains a dense family
  of scaling-periodic solutions. Furthermore, 
there are eternal solutions with 
trajectories {\em dense} in the scaling attractor --- 
we call these {\it Doeblin solutions}. 
And, for any given scaling trajectory, there is a dense set of initial data
whose forward trajectories shadow the given one.
\eit

In addition, this study of scaling limits reveals how Smoluchowski dynamics 
``compactifies'' in a natural way that accounts for clusters 
of zero and infinite size (dust and gel). 
Considering defective limits on $(0,\infty)$ that concentrate probability 
at 0 and $\infty$ yields a well-posed dynamics of 
``extended solutions'' on $[0,\infty]$. 
Proper solutions remain fundamental, 
but considering extended solutions with dust and gel helps to understand
just how the tails of initial data determine long-time behavior.

We remark that scaling-periodic solutions are analogous to the {\it
semi-stable} laws in probability theory \cite{semistable}.  
Our ``Doeblin solutions'' are constructed by ``packing the tails'' of the
\Levy\ measure in a fashion entirely analogous to the construction of
Doeblin's universal laws in probability \cite[XVII.9]{Feller}.
In this connection, it is interesting to note
that the examples in~\cite[XVII.9]{Feller}, dismissed by Feller as
``primarily of curiosity value,'' closely resemble
modern treatments of chaos, and Doeblin's construction appears
particularly prescient. 

The solvable cases of Smoluchowski's
equation correspond to sophisticated stochastic models with a rich
theory (see~\cite{Aldous, B_icm} for excellent reviews), 
so perhaps it is no accident that the
classical probabilistic  methods work so well. But let
us stress that our work really relies only on the analytical methods
for studying scaling limits that lie behind the classical limit theorems. 
These methods are simple and powerful and should be of utility for
understanding scaling phenomena in other applications that have no
obvious probabilistic meaning.  
Thus, no knowledge of probability is presumed and (almost) all details are
included  (though there is no substitute for reading Feller!).

\section{Statement of results}
\label{sec:results}
In this section, we state our results precisely in a 
setting that unifies the treatment of dynamic scaling for all the
solvable kernels. 

Let $E$ denote the open interval $(0,\infty)$, 
$\M$ the space of non-negative Radon measures on $E$, and $\PP$ the space of
probability measures on $E$. We will always use the weak topology on
$\M$ and $\PP$. 
We also let $\Ebar$ denote the closed half line with point at
infinity, $\Ebar=[0,\infty]=[0,\infty)\cup\{\infty\}$, 
and let $\bar\PP$ be the space of probability measures on $\Ebar$.

Rigorous theories for solutions where
$\nu_t(dx)=n(t,x)\,dx$ is a general size-distribution measure on
$E=(0,\infty)$, thus accounting for both continuous and discrete
size distributions in one general setting, are based on the moment identity 
\begin{equation}\label{R.moment}
\frac{d}{dt}\int_E f(x)\nu_t(dx) = 
 \ \frac12\int_E\int_E
\tilde{f}(x,y) K(x,y)\nu_t(dx)\nu_t(dy),
\end{equation}
where $f$ is a suitable test function and $\tilde
f(x,y)=f(x+y)-f(x)-f(y)$, see \cite{MP1,Norris,FL-wellp}.  
Let $m_\theta(t) := \int_E x^\theta \nu_t(dx)$ denote the $\theta$-th
moment of $\nu_t$. By the results of \cite{MP1}, 
for a solvable kernel of homogeneity $\gamma$, 
any initial measure $\nu_{t_0}$ with finite $\gamma$-th moment 
$m_\gamma(t_0)$ determines a unique continuous weak solution
\be 
\la
{R.smol}
\nudot = (\nu_t(dx), t \in [t_0,\Tmax) ).
\ee
For convenience we can always scale the initial data so that 
\be
\label{R.normal2}
t_0:= \left\{ \begin{array}{rl} 1 &  (K=2),\\ 0 & (K=x+y),
\\ -1 &  (K=xy),   \end{array} \right.
\quad\mbox{and }\ %
m_\gamma(t_0)= \int_E x^\gamma \nu_{t_0}(dx) =1.
\ee
Then
$\Tmax=\infty$ for $K=2$ and $x+y$, and $\Tmax=0$ for $K=xy$.
%$\Tgel = t_0 + m_2(t_0)^{-1}$ \footnote{[[=??in terms of $m_3$]]}
%(the gelation time) for $K=xy$~\cite[\S 2]{MP1}.  
For each solvable kernel, 
$m_\gamma(t)=\int_Ey^\gamma\nu_t(dy)$ is an explicitly known 
function --- from \qref{R.moment} with $f(x)=x^\gamma$ we find
\begin{equation}\label{R.mgamma}
m_\gamma(t)= \left\{
\begin{array}{ll}
t^{-1} &(K=2),\\
1      &(K=x+y),\\ 
|t|^{-1} &(K=xy).
\end{array} \right.
\end{equation}

The solution $\nu_t$ is typically not a
probability measure because the total number of clusters decreases in
time, but there is a naturally associated
probability measure $F_t(dx)$ with distribution function $F_t(x)$ defined by
%\footnote{Is it supposed to be $(0,x)$ instead of $(0,x]$?? GM: The
%normalization $(0,x]$ is used by Feller, and fits better with the S-measures}
\begin{equation}\label{R.Fp}
%\mu_t((0,x))=
F_t(x) = \int_{(0,x]} y^\gamma \nu_t(dy) \left\slash
\int_E y^\gamma\nu_t(dy). \right.
\end{equation}
In this way, we regard Smoluchowski's equation 
as defining a continuous dynamical system on the phase space $\PP$. 

\subsection{Eternal solutions}
For exceptional initial data $\nu_{t_0}$ we may also solve backwards
in time (meaning $\nu_{t_0}$ is {\em divisible\/} under clustering
dynamics). The maximum possible interval of existence that can be
obtained in this way is denoted $(\Tmin, \Tmax)$, where $\Tmin,\Tmax$
depend only on the kernel and $\int_E x^\gamma  \nu_{t_0}(dx)$.  With
the normalization \qref{R.normal2}, the maximum possible interval of
existence turns out to be 
\be
\label{R.normal3}
(\Tmin,\Tmax) = \left\{ \begin{array}{ll} (0,\infty) &  (K=2), \\
  (-\infty, \infty)  &  (K=x+y), \\
(-\infty,0) & (K=xy). \end{array} \right.
\ee
Solutions which are defined on this maximum interval of existence 
are the analog of infinitely divisible laws in probability.  
\begin{defn}
\label{R.eternal}
A solution to Smoluchowski's equation that is defined for all
$t\in(\Tmin,\Tmax)$ is called an {\em eternal solution}.  
\end{defn}

\subsection{The scaling attractor}
A central idea in dynamical systems theory is to understand the 
long-time behavior of solutions through the notions of
attractor and $\omega$-limit sets. Coagulation equations transport mass
from small to large scales, and all mass escapes as $t \to \Tmax$.
To obtain non-trivial  long-time behavior we must rescale solutions. We
adopt the following definitions for such {\em scaling dynamical
  systems\/}. Below, $T_n \in [t_0, \Tmax)$, $\betarsc_n >0$.
  We will often use the same letter to denote a measure
and its distribution function, e.g., $F(x)=\int_{[0,x]}F(dx)$. 

\begin{defn}\label{R.omegadef}
The (proper) {\em scaling $\omega$-limit set} of a solution $\nu$ to
Smoluchowski's equation  is the set of
probability measures $\hat F$ on $E$ for which there exist sequences
$T_n\to\Tmax$, $\betarsc_n \to \infty$, such that $F_{T_n}(\betarsc_nx) \to
\hat F(x)$ at every point of continuity of $\hat F$. 
\end{defn}

\begin{defn}\label{R.attractor}
The (proper) {\em scaling attractor}, $\attr$, is the set of probability
measures $\hat F$ on $E$ for which there exists
a sequence of solutions $\nudot\nsup$ defined for $t \in [t_0,\Tmax)$, and
sequences $T_n \to \Tmax$,  $\betarsc_n\to\infty$,  such that
$F\nsup_{T_n}(\betarsc_nx) \to  \hat F(x)$
at every point of continuity of $\hat F$. 
\end{defn}

As a consequence of continuous dependence of solutions on
initial data (forward and backward in time), we will show that
the scaling attractor is an invariant set, and that
points on the proper scaling attractor and eternal solutions are
in one-to-one correspondence.

\begin{thm}\label{RT.AE}
\begin{enumerate}
\item[(a)]
The proper scaling attractor $\attr$ is invariant: If $\nu$ is a solution
of Smoluchowski's equation,
and $F_t\in\attr$ for some $t$, then the solution is eternal and
$F_t\in\attr$ for all $t\in(\Tmin,\Tmax)$.
\item[(b)]
A probability measure $\hat F$ belongs to 
$\attr$ if and only if $\hat F(dx) = x^\gamma \nu_{t_0}(dx)$ 
for some eternal solution $\nudot$, where $t_0$ is as in
\qref{R.normal2}. 
\end{enumerate}
\end{thm}

The perfect definition of an attractor remains elusive (see for
example, the discussion in~\cite[Ch. 1.6]{GH}).  
Definition~\ref{R.attractor} is perhaps the simplest 
for dynamical systems. It also has the virtue of
generalizing the probabilistic notion of domains of partial
attraction~\cite[XVII]{Feller}. However, some typical properties 
that hold in finite-dimensional dynamical systems do not hold here.
For example, it need not be the case that every solution has a non-empty
scaling $\omega$-limit set. Nor is $\attr$ closed. Defective limits are
possible, as shown in \cite{MP1}. See \cite{JainOrey} for a
discussion of related issues in the probabilistic context.
It turns out that we can cure these defects and account for limits
that involve mass concentrating at zero or leaking to infinity,  by
the simple expedient of allowing limits to be probability measures on
$\Ebar=[0,\infty]$.

\begin{defn}\label{R.full}
The full {\em scaling $\omega$-limit set} of a solution $\nu$ to
Smoluchowski's equation is the set of probability measures $\hat F$ on 
$\Ebar$ with the property in Definition~\ref{R.omegadef}.
The full {\em scaling attractor}, $\attrx$, is the set of
probability measures $\hat F$ on $\Ebar$ with the property in
Definition~\ref{R.attractor}.
\end{defn}

The space $\bar\PP$ of probability measures on $\Ebar$, 
equipped with the weak topology, is compact --- any 
sequence contains a converging subsequence.
We will show that Smoluchowski dynamics naturally extends
by continuity from $\PP$ to $\bar\PP$. Such ``extended solutions''
have probability
distributions that may include atoms at $0$ and $\infty$, allowing for the
possibility that clusters have zero size (``dust'') or
infinite size (``gel'') with positive probability.  
To interpret these physically, one
should recognize that of course $0$ and $\infty$ are idealizations
relative to a given scale of measuring cluster size. 

We defer detailed discussion of extended solutions to
section~\ref{S.extend}. There we extend
Theorem~\ref{RT.AE} to relate the full scaling attractor $\attrx$
(now a compact set that is the closure of $\attr$) to the set of
eternal extended solutions.  
Also, the \LK\ representation and linearization theorems in the next two
sections have elegant extensions involving extended solutions. 
%This representation is used to establish a basic property: $\attrx$ is the
%closure of $\attr$.  
First, however, we think it appropriate to focus on standard weak
solutions, and develop the theory without dust in our eyes, so to speak.

\subsection{\LK\/ representations}

In probability theory, infinitely divisible distributions
are parametrized by the \LK\/ representation theorem, which expresses 
the log of the characteristic function (Fourier transform) in terms of 
a measure that satisfies certain finiteness conditions.
%In particular~\cite[XIII.7]{Feller}, a probability measure 
%$F$ supported on $\ebar$ is infinitely divisible if and only
%if its Laplace transform has the form 
%$\intR \exp(-qx) F(dx)=\exp(-\Phi(q))$ where
%the Laplace exponent $\Phi$ admits the representation
In particular~\cite[XIII.7]{Feller}, a function $\omega(q)$
is the Laplace transform $\intR \rme^{-qx} F(dx)$
of an infinitely divisible probability measure 
$F$ supported on $\ebar$ if and only if
$\omega(q)=\exp(-\Phi(q))$ where
the Laplace exponent $\Phi$ admits the representation
\be
\label{R.LK2a}
\Phi (q) = \int_{[0,\infty)} \frac{1-\rme^{-qx}}{x} \sm(dx)
\ee
for some measure $\sm$ on $[0,\infty)$ that satisfies
\be
\label{R.gendef}
\int_{[0,x]} G(dy)<\infty \quad\mbox{and}\quad
\int_{[x,\infty)} y^{-1}\sm(dy)<\infty \quad\mbox{for all $x>0$.}
\ee
Equivalently, 
\begin{equation}\label{R.gendef2}
\int_{[0,\infty)} (1 \wedge y^{-1}) \sm(dy) <\infty.
\end{equation}
We need a name for measures with this property, 
although none seems standard.
Such measures determine the jump-size distribution for 
{\em subordinators} --- increasing continuous-time random walks with
stationary independent increments --- hence we call them 
{\em \smeass}. To handle defective limits, it is
convenient to allow $y\inv G(dy)$ to have an atom at $\infty$.
%\footnote{[[Compare Feller's canonical measures? ]]}
\begin{defn}
\label{R.candef}
A measure $G$ on $[0,\infty)$ is an {\em \smeas}\ if 
\qref{R.gendef2} holds.
A pair $(G,g_\infty)$ is called an {\em \barsmeas}\ on $\Ebar=[0,\infty]$ if 
$G$ is an \smeas\ and $g_\infty\ge0$.
$g_\infty$ is called the charge of $y^{-1}G(dy)$ at $\infty$,
and we will abuse notation by
denoting the pair $(\sm,g_\infty)$ by $G$.
In addition, we say that an \smeas\ (or \barsmeas) $\sm$ 
is {\em divergent\/} if
\be
\label{R.candef5}
\sm(0)>0 \quad\mbox{or}\quad 
\int_{E} y^{-1} \sm(dy) = \infty.
\ee
\end{defn}
\noindent

Here recall we use the notation $\sm(x)=\int_{[0,x]} \sm(dy)$.
If $g_\infty=0$ we identify $G$ with $(G,0)$.
The space of \barsmeass\ has a natural weak topology which
will prove fundamental in our study of scaling dynamics.
\begin{defn}
\la{R.proper}
A sequence of \barsmeass\ $\sm\nsup$ {\em converges} to an 
\barsmeas\ $\sm$ as $n\to\infty$, if at every point $x\in(0,\infty)$
of continuity of $\sm$ we have 
\begin{equation}\label{R.candef1}
\int_{[0,x]} \sm\nsup(dy)\to \int_{[0,x]} \sm(dy) 
\end{equation}
and 
\begin{equation}\label{R.candef2}
\int_{[x,\infty]} y\inv \sm\nsup(dy)\to \int_{[x,\infty]} y\inv \sm(dy) .
\end{equation}
\end{defn}
The integrals in \qref{R.candef2} include the charge at $\infty$, if
any.  We note that in view of the weak convergence implied by
\qref{R.candef1}, 
convergence of $\sm\nsup$ to an \smeas\ $\sm$ (having $g_\infty=0$)
is equivalent to \qref{R.candef1} together with the tightness condition 
\begin{equation}\label{R.candef3}
\int_{[x,\infty]} y\inv \sm\nsup(dy) \to 0 \quad\mbox{as $x\to\infty$, 
uniformly in $n$}.
\end{equation}

Bertoin's main theorem in \cite{B_eternal} shows that eternal solutions for
$K=x+y$ are in one-to-one correspondence with divergent \smeass. 
(More precisely, Bertoin formulates his result in terms of ``\Levy\/ pairs,'' 
separating the atom at the origin from a jump measure on $(0,\infty)$).  
We extend this result as follows.
Let $\nu$ be an arbitrary solution to Smoluchowski's equation 
for a solvable kernel of homogeneity $\gamma$. 
Since the $\gamma$-th moment of $\nu_t$ is finite, 
$x^{\gamma+1} \nu_t(dx)$ is an \smeas. 
Rescaling, we associate with $\nu_t$ 
the \smeas\ $\sm_t$ defined by
\be
\label{R.cansoln1}
\sm_t(dx) = x^{\gamma+1} \nu_t( \lambda(t) dx),
%\ee
%\quad\mbox{where}
\qquad %$\lambda(t), t \in (\Tmin,\Tmax)$ is defined by
%\be
%\label{R.cansoln2}
\lambda(t) = \left\{ \begin{array}{rl} 1 & (K=2),\\ \rme^t &
  (K=x+y),\\ |t|^{-1} & (K=xy). \end{array} \right.
\ee 
Our choice of rescaling ensures that if the total measure $\sm_t(E)$ is
finite for some $t$, then it is constant:
$\sm_t(E) = m_{\gamma+1}(t)/\lambda(t)^{\gamma+1}=$const.

One computes that
\be
\label{R.Sgro}
\intR y^{-1} \sm_t(dy) = 
\frac{m_\gamma(t) }{ \lambda(t)^{\gamma}} = 
\left\{ \begin{array}{rl} 
t^{-1} & (K=2),\\ \rme^{-t} & (K=x+y), \\ |t| & (K=xy)
\end{array} \right.
\ee
The well-posedness theorem \cite{MP1} implies that solutions of Smoluchowski's
equation normalized according to \qref{R.normal2} that exist
on any time interval $[t,\Tmax)$ 
{\em are in one-to-one correspondence} with \smeass\ $\hat\sm$ that
satisfy 
\[
\hat\sm(0)=0\quad\mbox{and}\quad
\intR y^{-1}\hat\sm(dy)= m_\gamma(t)/ \lambda(t)^{\gamma} ,
\]
via $\hat\sm=\sm_{t}$. Through studying the limit $t\dnto\Tmin$,
we find that eternal solutions may be characterized as follows.
\begin{thm}
\label{RT.LK}
\begin{enumerate}
\item[(a)]Let $\nu$ be an eternal solution for Smoluchowski's equation with
$K=2,x+y$ or $xy$. Then there is a divergent \smeas\ $\dsm$ 
such that 
$\sm_t$ converges to $\dsm$ as $t \dnto \Tmin$. 
\item[(b)] Conversely, for every divergent \smeas\ $\dsm$, 
there is a unique eternal solution $\nu$ 
  such that $\sm_t$ converges to $\dsm$ as $t \dnto \Tmin$.  
\item[(c)] 
Let $\Hmap\colon\attr\to\Mcan$
map the (proper) scaling attractor $\attr$
to the set $\Mcan$ of divergent \smeass\
by $\Hmap(\hat F)=\dsm$, where $\dsm$ is the divergent \smeas\ 
associated to the eternal solution $\nu$ such that
$\hat F(dx)=x^\gamma \nu_{t_0}(dx)$ with $t_0$ as in
\qref{R.normal2}. Then $\Hmap$ is a bi-continuous bijection.
%\footnote{GM: Is this OK? The attractor isn't
% closed, but the statement seems true}
%[[?? Let $\sm$ be divergent, $\tilde \sm$ not, and $a_n\dnto0$, then
%$a_n\sm+\tilde \sm$ converges to $\tilde \sm$. What gives?]]
% Ah, neither $\Mcan$ nor $\attr$ is closed.
\end{enumerate}
\end{thm}
The procedure for obtaining the eternal solution $\nu$ from the 
divergent \smeas\ $\dsm$ is nonlinear and is different
for each kernel (see Theorems~\ref{CT.eternal},~\ref{AT.eternal}, and
\ref{MT.eternal} below).  It seems natural to call Theorem~\ref{RT.LK} a
\LK\/ representation for the scaling attractor $\attr$ --- as we will see,
eternal solutions are determined through the Laplace exponents
associated with divergent \smeass. 

In section~\ref{S.extend}, the correspondence in Theorem~\ref{RT.LK}
is expanded to one between eternal extended solutions and arbitrary
\barsmeass.

%The divergence condition \qref{R.candef5} seems peculiar to 
%Smoluchowski's equation and has important probabilistic meaning in
%terms of the smoothness of sample paths of the associated \Levy\/ process.

\subsection{Linearization of ultimate dynamics}
There are two natural group actions on the class of
eternal solutions that are related to scaling dynamics, 
arising from {\em time evolution} and {\em rescaling of size}.
A straightforward but remarkable consequence of the scaling properties
of Smoluchowski's equation is that nonlinear dynamics (time evolution)
on the scaling attractor $\attr$ is conjugate to a simple linear
scaling transformation of the divergent \smeass\ that correspond by
Theorem~\ref{RT.LK}.

\begin{thm}\label{RT.scale}
Let $\nu$ be a solution of Smoluchowski's equation with
$K=2$, $x+y$ or $xy$. Given scaling parameters $a>0$ and $b>0$, let
\begin{equation}\label{newnu}
\tnu_t(dx) = \left\{
\begin{array}{ll}
a \nu_{at}(b\,dx)       &(K=2),\\
b \nu_{t+\log a}(b\,dx) &(K=x+y), \\
ab^2 \nu_{at}(b\,dx)    &(K=xy),
\end{array} \right.
\end{equation}
with associated probability distribution function
\begin{equation}\label{newF}
\tilde F_t(x) = \left\{
\begin{array}{ll}
F_{at}(bx) &(K=2 \text{ or }xy),\\
F_{t+\log a}(bx) &(K=x+y). 
\end{array}\right.
\end{equation}
Then $\tnu$ is again a solution.
% on the interval $[\tilde{t_0},
%  \Tmax)$ where [[ugh??]] %$\tilde{t_0}$ is defined by
%\begin{equation}\label{newt0}
%\tilde{t}_0 = \left\{
%\begin{array}{ll}
%a^{-1} t_0 &(K=2 \text{ or }xy),\\
%t_0 - \log a &(K=x+y). 
%\end{array}\right.
%\end{equation}
If $\nu$ is eternal and $\dsm$ its associated divergent \smeas,
then $\tnu$ is eternal and its associated divergent \smeas\ is given by 
\begin{equation}\label{newdmg}
\tilde\dsm(x) = \left\{
\begin{array}{ll}
ab^{-1}\dsm(bx)         &(K=2),\\
a^2b^{-1}\dsm(a^{-1}bx) &(K=x+y), \\
a^{-2}b^{-1}\dsm(abx)   &(K=xy).
\end{array} \right.
\end{equation}
\end{thm}

\begin{proof} The proof is simple, based on Theorem~\ref{RT.LK} and 
the scaling properties of Smoluchowski's equation. First, one
checks that \qref{newnu} determines a solution, by
scaling the moment identity \qref{R.moment} in each case.
Next, compute that if the \smeas\ $\sm_t$ is associated with
$\nu_t$ as in \qref{R.cansoln1}, then the corresponding 
\smeas\ associated with $\tnu_t$ is given by 
\begin{equation}\label{newmg}
\tilde\sm_t(dx) = \left\{
\begin{array}{rll}
x\tnu_t(dx)        &=\ ab^{-1}\sm_{at}(b\,dx)         &(K=2),\\
x^2\tnu_t(\rme^t\,dx) &=\ a^2b^{-1}\sm_{t+\log a}(a^{-1}b\,dx) &(K=x+y), \\
x^3\tnu_t(|t|^{-1}dx) &=\ a^{-2}b^{-1}\sm_{at}(ab\,dx)   &(K=xy).
\end{array} \right.
\end{equation}
Then take $t\downarrow\Tmin$ and apply Theorem~\ref{RT.LK} to
deduce \qref{newdmg}.
\end{proof}

\begin{thm}\label{RT.time}
Let $\nu$ be an eternal solution with corresponding
divergent \smeas\ $\dsm$ and let $F_t$ be as in
\qref{R.Fp} for $K=2$, $x+y$ or $xy$. For each
$t\in(\Tmin,\Tmax)$, let $\dsm_t=\Hmap(F_t)$
be the divergent \smeas\ associated to $F_t\in\attr$.
Then
\begin{equation}\label{Ht}
\dsm_t(x) = \left\{
\begin{array}{ll}
t \dsm(x)            &(K=2),\\
\rme^{2t} \dsm(\rme^{-t}x) &(K=x+y),\\
|t|^{-2}\dsm(|t|x)   &(K=xy).
\end{array}\right.
\end{equation}
\end{thm}

\begin{proof} Take $b=1$ and put $t=t_0$ in \qref{newF}, then substitute
$a=t$, $\rme^t$, $|t|$ for $K=2$, $x+y$, $xy$ respectively to obtain
$\tilde F_{t_0}=F_t$. Then the corresponding divergent 
\smeas\ $\dsm=\Hmap(\tilde F_{t_0})$ is found from \qref{newdmg}.
\end{proof}

By this theorem, we see that in terms of the divergent
\smeas\ that corresponds to the solution, the time evolution
on the scaling attractor $\attr$ is governed by the linear equations
\begin{alignat}{2}
t \D_t H_t &= H_t  &\quad&(K=2),\\
(\D_t + x\D_x) H_t &= 2H_t &&(K=x+y), \\
(t \D_t - x\D_x) H_t &= -2H_t &&(K=xy).
\end{alignat}
%% Though simple, these equations are striking from a dynamical
%% systems viewpoint.  One combines these dynamics with an
%% arbitrary time-dependent size rescaling in order to study
%% scaling limits.  (Size rescaling and time evolution commute.)
%% This means that the study of scaling limits on the attractor 
%% is reduced to the study of a continuous dilation map 
%% with arbitrary choice of amplitude normalization.
%[[Must discuss why this indicates chaos, where the shift map is.]]
%(Dilation of $x$ by the factor $b$ corresponds to 
%a shift by $\log b$ after a logarithmic
%change of variables $x \mapsto \log x$.)
\subsection{How initial tails encode scaling limits}

The long-time scaling behavior is very sensitive to the initial
distribution of the largest clusters in the system, as indicated by
the characterization of domains of attraction via
\qref{1.domain}, and the linearization theorem above. 
In fact, the long-time scaling dynamics is encoded in the tails
of initial data in a simple fashion related to the \LK\
representation. 
%  In this section we describe two manifestations of this:
% \begin{itemize}
% \item[(i)] For any point $\hat F$ in the proper scaling attractor $\attr$,
% its associated divergent \smeas\ $H$ is a limit of
% scaled initial tails of the solutions that approximate $\hat F$.
% \end{itemize}

%In what follows,
%we consider sequences $a_n \to \Tmax, b_n \to \infty$ and rescaled
%solutions $\tilde{F}\nsup_t$ and 
%$\tilde{G}\nsup_t$ defined on $[\tilde{t}\nsup_0, \Tmax)$ as in
%\qref{newdmg}, \qref{newt0} and \qref{newmg} with 
%$a,b$ replaced by $a_n,b_n$. Observe that $\tilde{t}\nsup_0 \to
%\Tmin$.  For brevity, we define
%\[ \tilde\sm\nsup(dx) = \tilde{\sm}_{\tilde{t}\nsup_0}(dx) =??.\] 

\begin{thm}
\label{RT.initial_tails}
Let $\hat F\in\attr$ 
%so $\hat F(dx)=x^\gamma\nu_{t_0}(dx)$ where $\nu$ is an eternal solution, 
with associated divergent \smeas\ $H$.
Let $\nu\nsup$ be any sequence of solutions defined for $t\ge t_0$,
with associated initial \smeass\ given by
$\sm\nsup(dx)=x^{\gamma+1}\nu\nsup_{t_0}(dx)$. 
Let $T_n \to \Tmax$, $\beta_n\to\infty$. 
Then the following are equivalent: 
\bit
\item[(i)] ${F}\nsup_{T_n}(\beta_n x) \to \hat F(x)$ % = x^\gamma\nu_{t_0}(dx)$
as $n\to\infty$, at every point of continuity. 
%(Equivalently,
%$a_n\nu\nsup_{a_n}(b_n\,\cdot)$ converges weakly to $\hat F$.)
%% \item[(ii)] 
%% As $n\to\infty$, $a_n \vp\nsup_1(q/b_n) \to \Vp(q)$ for all $q\ge 0$. 
\item[(ii)] The rescaled initial \smeass\ $\tilde\sm\nsup$ defined by
\begin{equation}\la{R.initG}
\tilde\sm\nsup(x) = \left\{
\begin{array}{ll}
\beta_n^{-1}T_n \,\sm\nsup(\beta_nx)
%= T_n\beta_n^{-1}\int_{[0,\beta_nx]} y\nu\nsup_{t_0}(dy) 
         &(K=2),\\[6pt]
\beta_n\inv \rme^{2T_n}\, \sm\nsup(\rme^{-T_n}\beta_nx) 
%= \beta_n\inv\rme^{2T_n} \int_0^{x\beta_n\exp(-T_n)}y^2\nu\nsup_0(dy) 
&(K=x+y), \\[6pt]
\beta_n\inv |T_n|^{-2} \,\sm\nsup(|T_n|\beta_n x)
 &(K=xy),
\end{array} \right.
\end{equation}
have the property that
$\tilde\sm\nsup$ converges to $\dsm$ as $n \to \infty$.
\eit
\end{thm}
\noindent
This result generalizes to the full attractor $\attrx$, with $H$
replaced by the corresponding \barsmeas, see section~\ref{S.extend}.
We remark that in the proof it is shown that for the convergence in
part (ii) to hold, it is necessary that $\rme^{-T_n}\beta_n\to\infty$
for $K=x+y$, and $|T_n|\beta_n\to\infty$ for $K=xy$.
%[[To see this, follow the proof of Theorem 7.1 on p 1224 of \cite{MP1}
%up to (7.8), considering only sequential limits. It's hard to see 
%how one could give a direct argument without a solution formula.]]

%% \begin{rem}
%% %We note that $\limsup a_n/b_n\le1$ [[why??]]. 
%% If the sequence $\nu\nsup$ is constant, then $\hat F$ belongs to the
%% scaling $\omega$-limit set of this solution. 
%% [remove?: If also $a_n=b_n$ then
%% $\tilde\sm\nsup$ is a pure dilation of the initial mass
%% measure.  So if the initial mass is finite then (ii) holds
%% with $\dsm$ an atom at the origin.  Correspondingly $\nu$ is
%% a self-similar solution with exponentially decaying tail.
%% The solution $\nu\nsup$ belongs to its domain of attraction;
%% see \cite[Theorem 5.1]{MP1}].
%% \end{rem}

\subsection{Signatures of chaos} \la{RS.chaos}

The dilational representation of dynamics in \qref{Ht}
in terms of the \LK\ representation 
means that Smoluchowski dynamics on the scaling attractor is
a continuous analog of a Bernoulli shift map, a classical paradigm for
chaotic dynamics. We demonstrate the utility of this representation by
constructing solutions with both chaotic and regular orbits, and 
by proving a shadowing theorem illustrating sensitive dependence on the tails.

%\medskip
%\noindent
%{\em Solutions with dense limit sets.}
\subsubsection{Solutions with dense limit sets}
\begin{thm}\label{RT.dense}
There exists an eternal solution $\nu$ whose scaling $\omega$-limit
set contains every element of the full scaling attractor $\attrx$.
\end{thm}
\noindent
We call such solutions {\em Doeblin  solutions\/} by analogy with
Doeblin's universal laws. The construction follows Feller closely 
and relies only on general principles (separability of $\PP$ and $\Mcan$, and
continuity of the bijection $\Hmap$). Theorem~\ref{RT.dense} 
tells us that $\attrx$ cannot be decomposed into 
invariant subsets. 
\medskip
%[[What about density of Doeblin solutions?]] -- done in shadowing section

\subsubsection{Scaling periodic solutions}
%\label{RS.periodic}

%\noindent
%{\em Scaling-periodic solutions.}
Another classical signature of chaos is the density of periodic
solutions. The notion of periodicity generalizes as follows.
\begin{defn}
\label{RD.periodic}
Let $\nu$ be a solution and define $F_t(x)$ by
\qref{R.Fp}. We say $\nu$ is  {\em scaling-periodic}
if for some $t_1>t_0$ and $\betarsc >1$, 
\begin{equation}\label{R.periodic}
F_{t_1}(\betarsc x)=F_{t_0}(x) \quad\mbox{for all $x>0$.}
\end{equation}
\end{defn}
\noindent
These are analogous to semi-stable laws in probability 
theory~\cite{semistable}. The
\LK\/ representation yields a simple characterization.
\begin{thm}
\label{RT.periodic}
A scaling-periodic solution of Smoluchowski's equation with kernel
$K=2, x+y$ or $xy$ is eternal, and its divergent \smeas\ $\dsm$ satisfies
\begin{equation}
\dsm(x)=a\dsm(bx)
\end{equation} for some $a>0$, $b>1$ such that either  
\begin{enumerate}
\item[(i)] $a=1$ and $\dsm$ is an atom at the origin, or
\item[(ii)] $a<1$  and $ab>1$.
%and $\dsm$ is determined by its restriction to $[1,b)$. 
\end{enumerate}
Conversely, if 
$\dsm$ is a measure on $[0,\infty)$ with $\dsm(x)=a\dsm(bx)$, 
where $a>0$, $b>1$ and (i) or (ii) hold,
then $\dsm$ is a divergent \smeas\ and
the corresponding eternal solution is scaling-periodic.
\end{thm}
Case (i) is simple but important. The corresponding scaling-periodic
solutions are the self-similar solutions with exponential decay. More
generally, all self-similar solutions are determined by divergent
\smeass\ of the power-law form $\dsm(x)= C_\rho
x^{1-\rho}$, $0 < \rho \leq 1$. Scaling-periodic solutions that
are not self-similar solutions are generated by (ii).  
Thus, there are uncountably many scaling-periodic solutions. 
Moreover, the \LK\/ representation allows us to prove 

%they are dense in the full attractor $\attrx$.
\begin{thm}
\label{RT.Pdense}
Scaling-periodic solutions are dense in $\attrx$.
\end{thm}

% \noindent
% 

\subsubsection{Shadowing and sensitive dependence on the initial tails}

%\medskip
%\noindent
%{\em Shadowing and sensitive dependence on the initial tails.\/}
We show that asymptotically similar initial tails imply  {\em
  shadowing} of scaled solution trajectories. 
To study shadowing, we note that the space $\bar\PP$ of 
probability measures on $\Ebar$ is metrizable and compact.
We let $\dist(\cdot,\cdot)$ denote any metric on $\bar\PP$ which
induces the weak topology.

\begin{thm} \label{RT.shad}
Let $\nu$ and $\bar\nu$ denote any two solutions of Smoluchowski's
equation defined on $[t_0,\Tmax)$, and let the associated initial
\smeass\ be
\begin{equation}\label{R.shad1}
\sm(dx)=x^{\gamma+1}\nu_{t_0}(dx), \qquad
\bar\sm(dx)=x^{\gamma+1}\bar\nu_{t_0}(dx),
\end{equation}
with Laplace exponents $\vp$ and $\bar\vp$ respectively associated
as in \qref{R.LK2a}.
Assume that 
\begin{equation}\label{R.shad2}
%{\bar\sm(x)}/{\sm(x)} \sim L(x) \quad\mbox{as $x\to\infty$},
\bar\vp(q)/\vp(q) \sim L(1/q) \quad\mbox{as $q\to0$,}
\end{equation}
where $L$ is slowly varying at $\infty$.  
Suppose that $b(t)\uparrow\infty$ as $t\uparrow\Tmax$, and define 
\begin{equation}\la{R.bartb}
(\bar t,\bar b) = 
%\bar t= 
\left\{
\begin{array}{ll}
(\ t/L(b),\ b\ )   & (K=2),\\[5pt]
(\ t - \log L(b\rme^{-t}),\ b/L(b\rme^{-t}) \ )  & (K=x+y), \\[5pt]
(\ t\,L(|t|b) ,\ b L(|t|b)\ ) & (K=xy),
\end{array}\right.
\ee
so that $\bar b/\lambda(\bar t)=b/\lambda(t)$ with
$\lambda(t)$ as in \qref{R.cansoln1}.
Then we have 
\begin{equation}\la{R.dto0}
\dist( F_t(b(t)\,dx),\bar F_{\bar t}(\bar b(t)\,dx)) \to 0 \qquad
\mbox{as $t\to\Tmax$.}
\end{equation} 
\end{thm}
The simplest situation, requiring no readjustment of the scaling
($\bar t=t$, $\bar b=b$) is when $L=1$. 
When condition \qref{R.shad2} holds, the solutions $\nu$ and $\bar\nu$
have identical scaling $\omega$-limit sets, for example.
If one of the solutions in this theorem, say $\bar \nu$, is
self-similar, then the sufficient condition \qref{R.shad2} for
shadowing in this theorem is equivalent to \qref{1.domain} 
(see \cite{MP1}, (5.3) and (5.7) for $K=2$, and (7.2)
and (7.4) for $K=x+y$).  Hence \qref{R.shad2} is also necessary,
according to the classification theorem on domains of attraction.  It
appears that in general the sufficient condition for shadowing given
in this theorem may not always be necessary. But we will not pursue
this issue here. 

The sensitivity of solutions to initial tails in the weak topology
is revealed strikingly in Theorem~\ref{RT.shad}. 
The topology of weak convergence is undoubtedly natural 
for limit theorems, for example the approach to self-similarity.
On the other hand, this topology cannot distinguish the tails, 
as the following ``cut-and-paste'' argument shows. 
Let $\hat F=x^\gamma\nu_{t_0}$ and $\Fcheck=x^\gamma\nucheck_{t_0}$ 
be given initial data for two solutions, and define
\begin{equation}
\hat F\nsup(x) = \left\{
\begin{array}{ll}
\hat F(x) \wedge \Fcheck(n) & x< n\\
\check F(x) & x\ge n.
\end{array}\right.
\end{equation}
Then $\hat F\nsup\to \hat F$ as $n\to\infty$, 
and $\hat F\nsup$ has Laplace exponent given by 
\be
\frac{\vp\nsup(q)}q =  \int_0^n 
\left( \frac{1-\rme^{-qy}}{qy}\right) y\hat F(dy)
+\int_n^\infty
\left( \frac{1-\rme^{-qy}}{qy}\right) y\check F(dy) ,
\ee
from which one sees easily that if $\check\sm(E)=\infty$, then
$\vp\nsup(q)\sim\check\vp(q)$ as $q\to0$. Thus,
according to Theorem \ref{RT.shad},
the solution $\nu\nsup$ generated by $\hat F\nsup$ will shadow
$\nucheck$. This justifies the statement made
in the introduction that for any given scaling trajectory, there is a
dense set of initial data whose forward trajectories shadow the given one.

\subsection{Plan of the paper}
In section 3 we establish some basic facts
regarding convergence of the measures that generate the \LK\
representation. The analysis of eternal solutions is different in
detail for the constant and additive kernels, so we treat these cases
in turn, establishing Theorems~\ref{RT.AE}, \ref{RT.LK}, and
\ref{RT.initial_tails} for these kernels in sections 4 and 5.  
The multiplicative case reduces mathematically to the
additive by a change of variables, and is treated briefly in section~6. 
We emphasize that the results in this case concern the behavior of
solutions approaching the gelation time, so perhaps this is the case
of most interest physically.  

With the \LK\ representation in hand, we then construct Doeblin
solutions in section 7 and scaling-periodic solutions in section 8.
Extended solutions and the full scaling attractor $\attrx$ are studied
in section 9, and the shadowing theorem~\ref{RT.shad} is proved in
section 10, where we also provide a streamlined treatment of
the domains of attraction.

\section{Laplace exponents and limits of \barsmeass}
\label{S.laplace}
The main analytic tool in the study of the solvable kernels is the
Laplace transform.  Recall that a sequence of probability measures
$F\nsup$ is said to converge weakly to a
probability measure $F$ if the distribution functions
$F\nsup(x) \to F(x)$ at every point of continuity of the limit.
It is basic~\cite[XIII.1]{Feller} that  $F\nsup$ converges
weakly to $F$ if and only if the Laplace transforms 
converge pointwise:
\[\int_E \rme^{-qx} F\nsup(dx) \to \int_E \rme^{-qx} F(dx), \quad 
\text{for all $q > 0$.}\]
We will need the following refinements of this result for \barsmeass. 
%The results are [[close to ones that are]]
%well-known, but  proofs are included for completeness.   
With any \barsmeas\ $\sm$ we associate ``Laplace
exponents'' $\Vp$ and $\Psi$ (with $\Vp=\D_q\Psi)$ defined for 
$q \in \C_+=\{\lambda\in\C:\Re\lambda>0\}$ by
\begin{equation}
\la{L.1}
\Vp(q)  = \int_{\Ebar} \frac{1-\rme^{-qx}}{x} \sm(dx),
\quad
\Psi(q) = \int_{\Ebar} \frac{\rme^{-qx}-1 +qx}{x^2} \sm(dx) .
\end{equation}
If $g_0$ and $g_\infty$ denote amplitudes of the atoms of 
the measure $(1\wedge y\inv)\sm(dy)$ at $0$ and $\infty$, respectively,
this means
\begin{eqnarray}\label{L.1a}
\Vp(q) &=& q g_0 + g_\infty + \int_{(0,\infty)} \frac{1-\rme^{-qx}}{x} \sm(dx),
\\ \label{L.1b}
\Psi(q) &=&   \frac12 q^2 g_0 + q g_\infty +  
\int_{(0,\infty)} \frac{\rme^{-qx}-1 +qx}{x^2} \sm(dx).
\end{eqnarray}
We use the terminology Laplace exponent in accordance with
probabilists' usage. If we need to distinguish the two types
of exponents, we will refer to $\Vp$ 
 and $\Psi$ as Laplace exponents of
the first and second order, respectively. Observe that
\be
\la{L.3}
\partial_q \Vp = \partial_q^2 \Psi = \int_{[0,\infty)}\rme^{-qx}\sm(dx).
\ee 
These functions are Laplace transforms of a positive measure, 
thus are completely monotone functions on $(0,\infty)$. 

We note that the amplitude of the atom of $(1\wedge y\inv)\sm(dy)$ 
at $\infty$ is 
\begin{equation} \label{R.ginf}
g_\infty = \lim_{q\to0^+}\Vp(q) = \lim_{q\to0^+} q\inv\Psi(q),
\end{equation}
thus the \barsmeas\ $\sm$ is an \smeas\ if and only if this vanishes.
Furthermore, we claim that $\sm$ is divergent if and only if 
\begin{equation} \label{R.divcon}
\lim_{q\to\infty} \Vp(q) = \infty, \qquad\mbox{equivalently}\quad
\lim_{q\to\infty} q\inv\Psi(q)=\infty.
\end{equation}
To prove this, 
observe that
\[ \Phi(q) \leq q g_0+\int_{(0,\infty)} x^{-1} \sm(dx).\]
Thus, if $\lim_{q \to \infty} \Phi(q)=\infty$, then $\sm$ satisfies
\qref{R.candef5}. Conversely, if $\sm$ satisfies \qref{R.candef5}, then 
$\lim_{q \to \infty} \Vp(q)=\infty$ by the monotone
convergence theorem. The proof for $\Psi$ is similar. We integrate by
parts and use Fubini's theorem to obtain
\begin{eqnarray*}
 q^{-1}\Psi(q) &=&
\frac12 q g_0 +g_\infty+ \int_{(0,\infty)} 
(1-\rme^{-qx}) \int_{(x,\infty)}y^{-2} \sm(dy)\, dx \\
&\leq &
\frac12 q g_0 +g_\infty+ \int_{(0,\infty)} y^{-1} \sm(dy). 
\end{eqnarray*}
Thus, if $q^{-1} \Psi(q) \to \infty$ then $\sm$ satisfies
\qref{R.candef5}. The converse follows from the monotone convergence
theorem.

\begin{thm}
\label{LT.Mconv}
Let $\sm\nsup$ be a sequence of \barsmeass\ with Laplace
exponents $\Vp\nsup$ and $\Psi\nsup$. Then, taking $n \to \infty$, the
following are equivalent: 
\begin{enumerate}
\item[(i)] $\sm\nsup$ converges to an \barsmeas\ $\sm$
with Laplace exponents $\Vp$ and $\Psi$. 
\item[(ii)]  $\Vp(q):= \lim_{n\to\infty}\Vp\nsup(q)$ exists for each $q>0$. 
\item[(iii)] $\Psi(q):= \lim_{n\to\infty}\Psi\nsup(q)$ exists for each $q>0$. 
  \end{enumerate}
\end{thm}

\begin{proof} %[Proof of Theorem~\ref{LT.Mconv}]
(i) implies (ii): Fix $q>0$ and let $\veps >0$. 
\qref{R.candef2} allows us to choose
$a$ such that $a$ is a point of continuity for $\sm$ and for every $n$ 
\[  
\int_{[a,\infty]} \rme^{-qx} x\inv (\sm\nsup(dx)+\sm(dx))
\le \rme^{-qa} C <\veps.
\]%\int_a^\infty (1-\rme^{-qx}) x^{-1}\sm\nsup(dx) < \veps.\]
On the other hand, \qref{R.candef1} guarantees 
\[\int_0^a  (1-\rme^{-qx})x^{-1} \sm\nsup(dx) \to \int_0^a
(1-\rme^{-qx})x^{-1} \sm(dx).\] 
Using again \qref{R.candef2}, 
we conclude that for large $n$, $|\Vp\nsup(q)-\Vp(q)|<\veps$.
%Therefore, $\limsup\Vp\nsup(q) -2\veps \leq \Vp(q) \leq
%\liminf\Vp\nsup(q) + 2\veps$. 

(ii) implies (i): {\em 1. Claim:\/} $\Vp$ is analytic in $\C_+$
and $\Vp\nsup \to \Vp$ uniformly on compact subsets of $\C_+$.
{\em Proof:\/} Let $K \subset \C_+$ be compact. The claim follows
from the estimate: 
\be
\la{L.can1}
\sup_n \sup_{q \in K} |\Vp\nsup (q)| < \infty.
\ee
Indeed, by Montel's theorem, (\ref{L.can1}) implies
$\{\Vp\nsup\}_{n=1}^\infty$ are a normal family of analytic 
functions (i.e., precompact in the uniform topology). Thus, every
subsequence has a further subsequence converging uniformly 
to an analytic function. Since every subsequence  converges
(pointwise) to $\Vp$, this implies $\Vp\nsup \to \Vp$ uniformly and
$\Vp$ is analytic.  It remains to
prove \qref{L.can1}. We integrate by parts to obtain
\[ q^{-1} \Vp\nsup(q) = \sm\nsup(0) +
\int_{E} \rme^{-qx} \left( \int_{[x,\infty]}
  y^{-1}\sm\nsup(dy) \right) dx.\]
Thus, for any $a >0$,
  $\sup_{\Re q > a} |q^{-1} \Vp\nsup(q)| \leq a^{-1} \Vp\nsup(a)$. 
Since $\Vp\nsup(q)$ converges for all $q > 0$, we have $\sup_n
\sup_{\Re q > a} |q^{-1} \Vp\nsup(q)| < \infty$. This proves
\qref{L.can1}.

{\em 2. \/} Cauchy's integral formula and the claim imply
$\partial^k_q \Vp\nsup \to \partial^k_q \Vp$ for every $k \in \N$. 
Since $\partial_q \Vp\nsup$ are completely monotone, so is the limit
$\partial_q\Vp$.  

{\em 3.\/}  Thus, 
$\partial_q \Vp = \int_{[0,\infty)} \rme^{-qx} \sm(dx)$ 
is the  Laplace transform of a measure $\sm$ on $[0,\infty)$. 
We integrate with respect to $q$ with
$g_\infty:=\Vp(0^+)$ defined to be the charge of $y\inv\sm(dy)$ at $\infty$, 
and use Tonelli's theorem to obtain (\ref{L.1}). Note that $\int_{[1,\infty]}
x^{-1} \sm(dx)<\infty$ because $\Vp(q)<\infty$ for each fixed $q$,
so $\sm$ is an \barsmeas.

{\em 4.\/} The convergence $\D_q \Vp\nsup \to \D_q \Vp$  is
equivalent to weak convergence  of $\sm\nsup$ to $\sm$ on $[0,\infty)$, 
meaning \qref{R.candef1} holds. 
This implies that for every point $x$ of continuity of $\sm$,
as $n\to\infty$ we have
\begin{equation}
\int_{[0,x]} (1-\rme^{-qy})y\inv \sm\nsup(dy) 
\to \int_{[0,x]} (1-\rme^{-qy})y\inv \sm(dy),
\end{equation}
and together with (ii) this yields that
$\int_{[x,\infty]}y\inv\sm\nsup(dy)$ is bounded and 
\[
\int_{[x,\infty]} \rme^{-qy} y\inv \sm\nsup(dy)
\to \int_{[x,\infty]} \rme^{-qy}y\inv \sm(dy) .
\]
From (ii) then follows \qref{R.candef2}. This proves $\sm\nsup\to\sm$.

(ii) implies (iii): This 
is due to $\Psi\nsup(q)=\int_0^q\Vp\nsup(s)\,ds$ and monotonicity.
%is step {\em 2\/} [[??]] above.

(iii) implies (ii): Since $\Psi\nsup(q)=\int_0^q\Vp\nsup(s)\,ds
\ge \frac12q\Vp\nsup(\frac12q)$, we find that \qref{L.can1} holds as
in step {\em 1\/} above. Then for every subsequence of $\Vp\nsup$ there is a
further subsequence that converges on compact sets of $\C_+$ to an analytic
limit $\Vp$. This limit is unique due to (iii), and (ii) follows.  (It follows
also that $\Psi\nsup \to \Psi$ uniformly on compact sets.)
\end{proof}

\section{The constant kernel}
\label{sec:const-ker}
In this section we study eternal solutions and the \LK\/ representation 
in particular
%present a detailed analysis of the scaling dynamics
for the constant kernel $K=2$. 
This kernel is technically easiest to deal
with, and the general framework is most transparent.  
Theorems~\ref{CT.attr}, \ref{CT.eternal} and
\ref{CT.Mconv2} are the main technical results and serve to
establish Theorems~\ref{RT.AE} and \ref{RT.LK} for this kernel. 
%With these results in hand, 
%we exploit the scaling properties of the \LK\/ representation to gain
%a detailed  understanding of the ultimate scaling dynamics in the
%following [[??]] section.

\subsection{Preliminaries}
Smoluchowski's equation with constant
kernel $K=2$ has a unique global solution in an appropriate weak sense
given any  initial size-distribution measure with finite zero-th
moment~\cite[\S 2]{MP1}.  
For convenience, we adopt the normalization in \qref{R.normal2}.
The moment identity \qref{R.moment} is valid for all
bounded continuous functions $f$ on $\bar{E}$, and taking
$f=1$ we find that the total number density of clusters is
$\nu_t(E)=t\inv$.  
Since $t \nu_t(E)=1$, 
we associate to each solution a probability distribution function
\begin{equation}\label{C.Fdef}
F_t(x) = \int_{(0,x)} \nu_t(dx)\left\slash \int_E \nu_t(dx)\right. 
= t \nu_t(x).
\end{equation}
We also introduce the \smeass 
\be
\label{C.cansoln1}
\sm_t(dx) = x \nu_t(dx),
\ee
and associated Laplace exponents
%[[How about $\Vp_t(q)$ or $\vp_t(q)$ instead of $\vp(t,q)$ ?]]
\begin{equation}\label{C.vpdef}
\vp(t,q) = \int_E (1-\rme^{-qx}) \nu_t(dx) = 
\int_{\Ebar} \frac{1-\rme^{-qx}}{x} \sm_t(dx).
\end{equation}
%and put $\vp_1(q)=\vp(1,q)$.
%Note that the Laplace transform of $F$ is
%\begin{equation}\label{C.Fvp}
%\int_E \rme^{-qx}F_t(dx) = 1-t\vp(t,q).
%\end{equation}

Notice that $q\mapsto\vp(t,q)$ is strictly increasing with 
$\vp(t,\infty)= \nu_t(E)=t^{-1}$, and  
$\partial_q \vp(t,q)$ is the Laplace transform of the
mass-distribution measure $x\,\nu_t(dx)$, so is completely monotone.
$\vp$ solves the simple equation
\begin{equation}
\label{eq:vp0}
\partial_t \vp = -\vp^2, 
\end{equation}
for which the solution at any time $t>0$ is determined from data
at time $t_0>0$ according to 
\begin{equation}
\label{eq:vp1}
\quad \vp(t,q) =
\frac{\vp(t_0,q)}{1+(t-t_0)\vp(t_0,q)},  \quad q \geq 0,\ t >0.
\end{equation}
Since $0 \leq \vp(t_0,q) <t_0\inv$, we see that given $F_{t_0}=t_0\nu_{t_0}$
an arbitrary probability measure, 
$\vp(t,q)$ is well-defined on the time-interval $(0,\infty)$.  
But for $0<t<t_0$, $\vp(t,q)$ may not have completely
monotone derivative, and thus may not define a (positive)
measure. The map $q\mapsto\D_q\vp(t,q)$ is completely monotone 
for all $t \in (0,\infty)$ if and only if $\nudot$ is an eternal solution.  

%which has the unique solution
%\begin{equation}
%\label{eq:vp1}
%\quad \vp(t,q) =
%\frac{\vp_1(q)}{1+(t-1)\vp_1(q)},  \quad q \geq 0,\ t >0.
%\end{equation}
%Since $0 \leq \vp_1(q) <1$, we see that for {\em any} probability
%measure $\nu_1$, $\vp(t,q)$ is well-defined on the time-interval
%$(0,\infty)$.  But for $0<t<1$, $\vp(t,q)$ may not have completely
%monotone derivative, and thus may not define a (positive)
%measure. The map $q\mapsto\vp(t,q)$ is completely monotone 
%for all $t \in (0,\infty)$ if and only if $\nudot$ is an eternal solution.  

Our study of convergence properties for solution sequences is based on
pointwise convergence properties of $\vp$, which are equivalent to 
%proper 
convergence properties of the \smeass\ $\sm_t$ 
according to the results of section~\ref{S.laplace}.
We begin by proving the continuous dependence of solutions on initial
data, based on the evident fact that $\vp(t,q)$ is a continuous function of
$\vp(t_0,q)$.

\begin{thm}\label{CT.cts}
(Continuous dependence on data.)
For Smoluchowski's equation with constant kernel $K=2$, let $t_0>0$
and let $\nu\nsup$ be a sequence of solutions defined for $t\ge t_0$.
\begin{enumerate}
\item[(a)] If $\nu\nsup_{t_0}$ converges weakly to a measure $\hat\nu_0$
with $\hat\nu_0(E)=t_0\inv$, then for every $t\ge t_0$ we have
that $\nu\nsup_t$ converges weakly to $\nu_t$, the time-$t$ solution
with initial data $\nu_{t_0}=\hat\nu_0$.
\item[(b)] For any $t\ge t_0$, if $\nu\nsup_{t}$ converges weakly to a
measure $\hat\nu$ with $\hat\nu(E) =t\inv$, then 
$\nu\nsup_{t_0}$ converges weakly to a measure $\hat\nu_0$ with
$\hat\nu_0(E)=t_0\inv$, and $\hat\nu=\nu_t$, the time-$t$ solution 
with initial data $\nu_{t_0} = \hat\nu_0$.
\end{enumerate}
\end{thm}

\begin{proof} We prove part (b); part (a) is similar.
Let $\sm\nsup_t(dx)=x\nu\nsup_t(dx)$ and $\hat\sm(dx)=x\hat\nu(dx)$,
and let $\vp\nsup(t,q)$ 
and $\hat\vp(q)$ be the associated Laplace exponents as in \qref{C.vpdef}.
The hypothesis is equivalent to saying 
that the \smeass\ $\sm\nsup_t(dx)$ converge 
%properly 
to a non-divergent \smeas\  $\hat\sm(dx)=x\hat\nu(dx)$ with $\int_E
x\inv\hat\sm(dx)=t\inv$. 
This is equivalent to the statement that for all $q>0$,
$\vp\nsup(t,q) \to \hat\vp(q)$ as $n\to\infty$, where
$\hat\vp(\infty)=t\inv$ and $\hat\vp(0^+)=0$.
Then it follows that 
\begin{equation}\label{C.tt0}
\vp\nsup(t_0,q) 
= \frac{\vp\nsup(t,q)}{1-(t-t_0)\vp\nsup(t,q)}
\to 
\hat\vp_0(q):= \frac{\hat\vp(q)}{1-(t-t_0)\hat\vp(q)}.
\end{equation}
Since $\vp_0(0^+)=0$ and $\hat\vp_0(\infty)=t_0\inv$, 
we conclude that $\hat\vp_0$ is the Laplace exponent for a 
measure $\hat\nu_0$ on $E$ with $\hat\nu_0(E)=t_0\inv$,
and that $\nu\nsup_{t_0}$ converges weakly to $\hat\nu_0$.
We compare \qref{C.tt0} with the explicit solution (\ref{eq:vp1}) to
see that $\hat\nu=\nu_t$,
where $\nudot$ is the solution on $[t_0,\infty)$ with initial data $\hat\nu_0$.
\end{proof}

\subsection{The scaling attractor and eternal solutions}
We are ready to prove Theorem~\ref{RT.AE} for the kernel $K=2$.  
First we consider part (b), the correspondence between the scaling attractor
and eternal solutions.

\begin{thm}\label{CT.attr}
A probability measure
$\hat F$  is an element of the scaling attractor for
Smoluchowski's equation with constant kernel $K=2$ if and only if 
$\hat F = \nu_1$ for some eternal solution $\nudot$.
\end{thm}
\begin{proof}
Let us first suppose that $\hat  F = \nu_1$ for some eternal solution
$\nudot$ and show that $\hat  F \in \attr$. 
Pick arbitrary sequences $T_n, \beta_n \to \infty$, and consider
the sequence of rescaled  eternal solutions
\be
\label{C.attr-rsc}
 \nu_t\nsup(dx) = \frac{1}{T_n} \nu_{t/T_n} \left(\beta_n^{-1}
dx\right), \quad t>0. 
\ee
Observe that $\nu_t\nsup(E) = t^{-1}$, therefore,
\[ F\nsup_{T_n}(\beta_n x) = \nu_{1}(x) = \hat F(x) \]
for every $x$. Thus, $\hat F\in\attr$ by Definition~\ref{R.attractor}.

Conversely, suppose $\hat F \in \attr$. We shall show that $\hat F =
\nu_1$ for 
some eternal solution $\nudot$. Let $\hat \vp$ denote the Laplace
exponent of $\hat F$, and  $\nudot\nsup,T_n, \beta_n$ be as in
Definition~\ref{R.attractor}. Consider the rescaled measures 
\[
\tilde\nu\nsup_t(dx)=T_n\nu\nsup_{tT_n}(\beta_n dx).
\]
This rescaling yields a solution that is defined for $t\ge 1/T_n$,
and by hypothesis we have that $\tilde\nu\nsup_1$ converges
weakly to $\hat F$. Then by Theorem~\ref{CT.cts}, for any $t>0$
we infer that $\tilde\nu\nsup_t$ converges weakly to $\nu_t$
where $\nu_t(E)=t\inv$ and $\nu$ is a solution with $\nu_1=\hat F$. 
The solution $\nu$ is eternal since it is defined 
for $t\ge t_0$ for every $t_0>0$.
\end{proof}

Let us now prove that $\attr$ is invariant (part (a) of Theorem~\ref{RT.AE}).
Suppose $\nu$ is a solution on some time interval $[t_1,\infty)$, 
normalized so $\nu_t(E)=t^{-1}$.
Suppose $F_{T}\in\attr$ for some $T\ge t_1$.
Replacing $\nu_t(dx)$ by $T\nu_{Tt}(dx)$, we may presume $T=1$ without loss 
of generality. By Theorem~\ref{CT.attr} above, $F_T=\tilde\nu_1$ for 
some eternal solution $\tilde\nu$. But then $\nu_t=\tilde\nu_t$
for all $t\ge t_1$, meaning that $\nu$ is (the restriction of) an
eternal solution. We obtain that $F_t\in\attr$ for all $t>0$ by
a similar scaling argument.

\subsection{\LK\/ representation of eternal solutions}

\begin{thm}
\label{CT.eternal}
\begin{enumerate}
\item[(a)] Let $\nu$ be an eternal solution to Smoluchowski's
equation with $K=2$. Then there is divergent \smeas\  $\dsm$
such that as $t\dnto 0$, the mass measure
$\sm_t(dx)=x\,\nu_t(dx)$ converges to $\dsm$.
%properly  

\item[(b)] Conversely, given any divergent \smeas\  $\dsm$ there
  is a unique eternal solution with the properties in part (a),
  defined for all $t\in(0,\infty)$ via
\begin{equation}
\label{C.eternal}
\vp(t,q) = \frac{\Vp(q)}{1+t\Vp(q)},
\qquad
\Vp(q) = \int_\Ebar \frac{1-\rme^{-qx}}{x}H(dx). 
\end{equation}
\end{enumerate}
\end{thm}

\begin{proof}
We first show (a). Immediately from the solution formula
(\ref{eq:vp1}), 
\be
\label{eq:lim_vp1}
\lim_{t\to0}\vp(t,q) = 
 \lim_{t \to 0} \frac{\vp(1,q)}{ 1+(t-1) \vp(1,q)} = 
\frac{\vp(1,q)}{1-\vp(1,q)} =: \Vp(q) 
\ee
exists for all $q>0$, with $\Vp(q) < \infty$,
$\Vp(0^+)=0$, and $\Vp(\infty)=\infty$. 
By Theorem~\ref{LT.Mconv}, $\sm_t$ converges to an \barsmeas\/ $\dsm$
with Laplace exponent $\Vp$, and $\dsm$ is a divergent \smeas\/ by
the criteria in \qref{R.ginf}-\qref{R.divcon}.

Let us now prove (b). Let $\dsm$ be a divergent \smeas\  with
Laplace exponent $\Vp$. By \qref{eq:lim_vp1}, any eternal solution with
the properties in part (a) must be determined by \qref{C.eternal}.
Observe that the function $q/(1+tq)$ has completely monotone derivative
for $t \in (0,\infty)$. It follows that $\partial_q \vp(t,q)$ is
completely monotone when  $\vp(t,q)$ is given by
(\ref{C.eternal})~\cite[XIII.4]{Feller}.   
Moreover, with $\nu_t$ determined from \qref{C.vpdef},
$\nu_t(E)= \vp(t,\infty)= t^{-1}$. Thus, $\nu_t$ is indeed an eternal solution.
\end{proof}

\begin{rem}
Observe that $\sm_t(E)= \int_E x \nu_t(dx)$ is finite 
for some $t\in(0,\infty)$ if and only if it is finite for all $t$.
However, it is not necessary that 
the mass be finite for a solution to be well-defined.
\end{rem}

Theorem~\ref{CT.eternal} establishes parts (a) and (b) of 
Theorem~\ref{RT.LK}. To establish part (c), we need to show that the map
$\nu_1\mapsto\dsm $ from $\attr$ to $\Mcan$ is a bi-continuous bijection.
\begin{thm}\label{CT.Mconv2}
Let $\nu\nsup$ be a sequence of eternal solutions with corresponding
divergent \smeass\ $\dsm\nsup$. 
Fix $t>0$. Then, taking $n\to\infty$, the following are equivalent:
\bit
\item[(i)] 
$\nu\nsup_t$ converges weakly to some
measure $\hat\nu$ with $\hat\nu(E)=t\inv$. 
\item[(ii)] 
$\dsm\nsup$ converges to some divergent \smeas\  $\dsm$. 
%properly
\eit
If either (equivalently both) of these conditions hold, then 
$\hat\nu=\nu_t$ for an eternal solution with divergent \smeas\  $\dsm$.
\end{thm}
\begin{proof} 
Assume (i), so $\nu\nsup_t$ converges to $\hat\nu$ with
$\hat\nu(E)=t^{-1}$.  Then $\sm\nsup_t(dx)=x\nu\nsup(dx)$
converges to $\hat\sm(dx)=x\hat\nu(dx)$ and
the associated Laplace exponents converge: $\vp\nsup(t,q)\to\hat\vp(q)$
for all $q>0$. Hence
\begin{equation} \label{C.Pconv}
\Vp\nsup(q) = \frac{\vp\nsup(t,q)}{1-t\vp\nsup(t,q)} \to 
\Vp(q):= \frac{\hat\vp(q)}{1-t\hat\vp(q)}, 
\end{equation}
as $n\to\infty$ for every $q>0$. 
Since $\hat\vp(0^+)=0$ and $t\hat\vp(q)\to1$ as $q\to\infty$, 
$\Vp(q) < \infty$ for every $q >0$, $\Vp(0^+)=0$, 
and $\lim_{q \to \infty}\Vp(q)=\infty$. By Theorem~\ref{LT.Mconv} 
and \qref{R.ginf}-\qref{R.divcon}, this proves (ii). 

We now show (ii) implies (i). Suppose the divergent \smeass\ $\dsm\nsup$
converge to a divergent \smeas\  $\dsm$. Then
Theorem~\ref{LT.Mconv} with \qref{R.ginf}-\qref{R.divcon}
implies $\Vp\nsup(q)\to\Vp(q)$ for every $q>0$, 
$\Vp(0^+)=0$, and $\Vp(q)\to\infty$ as $q\to\infty$. Then, 
\[
\vp\nsup(t,q)=\frac{\Vp\nsup(q)}{1+t\Vp\nsup(q)}
\to \frac{\Vp(q)}{1+t\Vp(q)} =\vp(t,q)
\]
for every $q>0$. This yields weak convergence of $\nu\nsup_t$ to
$\nu_t$, where $\nu$ is the eternal solution with Laplace exponent
$\Vp$ and divergent \smeas\  $\dsm$.
\end{proof}

\subsection{Scaling limits and initial tails}
We now prove Theorem~\ref{RT.initial_tails} for the constant
kernel. 
\begin{proof}[Proof of Theorem~\ref{RT.initial_tails}]
Introduce rescaled solutions
$\tilde\nu\nsup_t(dx)=T_n\nu\nsup_{tT_n}(\beta_ndx)$,
and let $\tilde F\nsup_t = t\tilde\nu\nsup_t$. 
Also let $\tilde\sm\nsup_t(dx)= x\tilde\nu\nsup_t(dx)$ and let
$\tilde\vp\nsup(t,q)$ be the associated Laplace exponent.
Let $H$ be the divergent \smeas\  corresponding to $\nu$ and
$\Vp$ its Laplace exponent, and let $\vp(q)$ be the Laplace exponent
of $\sm(dx)= x\nu_1(dx)$.

Then statement (i) of Theorem~\ref{RT.initial_tails} is equivalent to saying $\tilde
F\nsup_1\to \hat F$ weakly, 
meaning the \smeass\/ $\tilde\sm\nsup_1$ converge to $\sm$
with $\int_E x\inv\sm(dx)=1$. This is equivalent to saying 
\begin{equation}\label{e.ph1}
\tilde\vp\nsup(1,q)\to \vp(q), \quad q>0, \quad 
\mbox{where $\vp(0^+)=0$,\ $\vp(\infty)=1$.}
\end{equation}
On the other hand, since $T_nx\nu\nsup_1(\beta_ndx)= \tilde\sm_{1/T_n}(dx)$, 
statement (ii) of Theorem~\ref{RT.initial_tails} is equivalent to the assertion
\begin{equation}\label{e.ph2}
\tilde\vp\nsup(T_n\inv,q) \to \Vp(q), \quad q>0, \quad
\mbox{where $\Vp(0^+)=0$,\ $\Vp(\infty)=\infty$.}
\end{equation}
But by the solution formulae \qref{eq:vp1} and \qref{C.eternal}, we have
\[
\tilde\vp\nsup(T_n\inv,q) = \frac{\tilde\vp\nsup(1,q)}
{1+(T_n\inv-1)\tilde\vp\nsup(1,q)},
\qquad
\Vp(q) = \frac{\vp(q)}{1-\vp(q)}.\]
Since evidently \qref{e.ph1} is equivalent to \qref{e.ph2}, (i)
is equivalent to (ii).
\end{proof}

\subsection{The representation at $+\infty$}

For the additive kernel, Bertoin showed that an eternal solution can
be uniquely identified by its asymptotic behavior as $t\to\infty$ also.  
For the constant kernel, an analogous result follows easily
from \qref{C.vpdef} and \qref{C.eternal}.

\begin{thm} Let $\nu$ be an eternal solution of Smoluchowski's
equation with constant kernel $K=2$, and let $\Vp$ be the Laplace
exponent of the divergent \smeas\  associated with $\nu$. 
Then as $t\to\infty$, 
the measure $t^2 \nu_t$ converges weakly on $(0,\infty)$ to 
a measure $\Lambda_+$ with Laplace transform
\[ \Vp_+(q):= \int_0^\infty \rme^{-qx}\Lambda_+(dx) = \frac{1}{\Vp(q)}
= \lim_{t\to\infty} t^2 \int_0^\infty \rme^{-qx}\nu_t(dx).
\]
\end{thm}

Clearly an eternal solution $\nu$ is uniquely determined from $\Lambda_+$ 
through $\Vp(q)=1/\Vp_+(q)$. 
We see that the measure $\Lambda_+$ has a Laplace transform
$\Vp_+(q)$ defined for all $q>0$, and $\Vp_+(q)\to\infty$ as
$q\to0$ since $\Vp(0^+)=0$. 
So $\int_0^1\Lambda_+(dx)<\infty$ and $\int_E\Lambda_+(dx)=\infty$.

The class of measures $\Lambda_+$ which arise in this way 
is characterized by the property that $\eta(q)=\D_q(1/\Vp_+(q))$ 
is the Laplace transform of some divergent \smeas\  $\dsm$
(i.e., $\eta$ is completely monotone, locally integrable on $[0,\infty)$
and $\int_E \eta(q)\,dq=\infty$).
There does not appear to be a simple characterization by moment
conditions. 

\begin{rem}
This representation has an interesting probabilistic interpretation;
see~\cite[p.74]{B_book}. 
If $X_\cdot$ is a subordinator with Laplace exponent $\Vp$, then
  $\Vp_+ = 1/\Vp$ is the Laplace transform of the potential measure $U$,
  defined on Borel sets $A \subset E$ by
$U(A) = \mathbb{E} \left( \int_0^\infty \mathbf{1}_{\{X_s \in A\}} ds
  \right)$.
\end{rem}

\section{The additive kernel}
\label{sec:add}
In this section we study the scaling dynamics for the additive kernel. 
Our main aims are to prove continuous dependence on initial data,
establish the correspondence between points on the scaling attractor
and eternal solutions, and revisit Bertoin's \LK\/ representation
with convergence of \smeass\ in mind.

\subsection{Solution by Laplace transform}
The solution of Smoluchowski's equation with kernel $K=x+y$ by the
Laplace transform is 
classical~\cite{Drake},  and remains the basis for rigorous
work. Let $t_0\in\R$ be arbitrary. 
We assume $\nu_{t_0}$ is a (possibly infinite) measure with
$\int_E x \nu_{t_0}(dx) < \infty$.  Without loss of generality,
we may assume $\int_E x\nu_{t_0}(dx)=1$.
We have shown~\cite[Thm 2.8]{MP1} that (\ref{eq:smol1}) has a unique
solution $\nu_t$ for $t \geq t_0$ in an appropriate weak sense, such that 
\be
\la{A.mass1}
\int_E x \nu_t(dx) = 1, \quad t \geq t_0.
\ee
As for the constant kernel, we use the notation
\begin{equation}\label{A.vpdef}
\vp(t,q) = \int_E (1-\rme^{-qx}) \nu_t(dx) , \quad q \geq 0.
%\qquad 
%\vp_1(q) = \int_E (1-\rme^{-qx}) \nu_1(dx),   
\end{equation}
and set $\vp_0(q)=\vp(t_0,q)$. 
To study scaling limits we  consider the mass distribution function,
which is the natural probability distribution function associated to a
solution. Let 
\begin{equation}\label{A.Fdef}
F_t(x) = \int_{(0,x]} y\, \nu_t(dy) .
\end{equation}
Note that the Laplace transform of $F_t$ is
\begin{equation}\label{A.Fvp}
\int_E \rme^{-qx}F_t(dx) = \partial_q \vp(t,q).
\end{equation}
Thus, $\partial_q \vp(t,q)$ is completely monotone and 
$\partial_q \vp(t,0)= 1$, $t \geq 0$. 
We know from \cite{MP1} that if we substitute 
$f(x)=1-\rme^{-qx}$ in (\ref{R.moment}) 
we find that $\vp(t,q)$ solves the hyperbolic equation
\begin{equation} 
\label{A.1}
\partial_t \vp -\vp \partial_q \vp = -\vp. 
\end{equation}

Following Bertoin, it is convenient to introduce the new variables
\begin{align}
\label{A.psi1}
&s = \rme^t, \quad s_0= \rme^{t_0},
%\\ \la{A.psi1}
%&\Vp(s,q)=\vp(t,q), 
\quad 
\psi(s,q) = \frac{q}{s} - \vp\left(t, \frac{q}{s}\right). 
\end{align}
By \qref{A.mass1} and \qref{A.vpdef}, $\psi$ is the
Laplace exponent 
\be
\la{A.psi2}
\psi(s,q) = \int_{E} y^{-2} \left(\rme^{-qy} - 1 + qy\right) 
\sm_{\log s}(dy), 
\ee
where $\sm_t$ denotes the \smeas\ 
\be
\la{A.psi3}
\sm_t(dx) = x^2 \nu_{t} (\rme^t \,dx).
\ee
Observe that $\sm_t$ is not a finite measure in general, but if
$\sm_{t}(E) <\infty$ for some $t$, then $\sm_t(E)$ is finite for 
for every $t$ for which the solution is defined, and is constant.
We substitute \qref{A.psi1} in \qref{A.1} to see that $\psi$ satisfies
the inviscid Burgers equation
\be
\la{A.psi4}
\partial_s \psi + \psi \partial_q \psi =0, \quad s > s_0.
\ee
The values of $\psi$, $\D_q\psi$ and $\D_q^2\psi$ are positive
for $s\ge s_0$, $q>0$, and $\D_q^2\psi(s,\cdot)$ is completely
monotone since it is the Laplace transform of $G_t$.
In addition, \qref{A.mass1}, \qref{A.psi2} and \qref{A.psi3} imply
\be
\la{A.psidivg}
\lim_{q \to \infty} \partial_q \psi(s,q) = \int_Ex^{-1}\sm_{\log s}
(dx) = s^{-1}. 
\ee

We may describe $\psi(s,q)$ globally for $s>s_0$ by the method of
characteristics.  A surprising fact is that we may always solve for
$\psi$ backwards in time, for all $s>0$, without developing singularities.
The solution need not correspond to a positive measure $\nu_t$
for $t <t_0$, however.  This is analogous to the situation for $K=2$.
\begin{lemma}
\label{AL.charsoln}
Let $t_0 \in \R$ and $\nu_{t_0} \in \M$ with $\int_E
x\,\nu_{t_0}(dx)=1$, and let $\psi_0(q_0)= q_0/s_0 - \vp_0(q_0/s_0)$. 
There is a unique solution $\psi(s,q)$ to \qref{A.psi4} defined for every 
$s >0$ and $q> 0$,  such that $\psi(s_0,\cdot)=\psi_0(\cdot)$.
\end{lemma}
\begin{proof} 
Applying the method of characteristics as usual, 
the solution $\psi=\psi(s,q)$ is determined implicitly from the equation
\begin{equation}\label{A.imp}
h(s,q,\psi):= \psi-\psi_0(q-(s-s_0)\psi)=0.
\end{equation}
We have $h(s,q,0)<0$, and $\D_\psi h>s/s_0$ since
$\D_q\psi_0<s_0^{-1}$ by \qref{A.psidivg}.
Since $\psi_0$ is analytic, \qref{A.imp} determines a solution
of \qref{A.psi4} analytic in $(s,q)$ for all $s>0$, $q>0$.
\end{proof}

Equation \qref{A.imp} determines the solution at time $s$ from data at time
$s_0$ and plays the same role in the analysis here as equation
\qref{eq:vp1} played in the previous section. Convergence properties
of solutions will be deduced from the pointwise convergence properties
of the Laplace exponent $\psi$ using the theory from section \ref{S.laplace}.

\begin{thm}\label{AT.cts} (Continuous dependence on data.)
For Smoluchowski's equation with additive kernel $K=x+y$, let $t_0\in \R$
and let $\nu\nsup$ be a sequence of solutions defined for $t\ge t_0$
with $\int_E x \nu\nsup_t(dx)=1$ for all $t \geq t_0.$
\begin{enumerate}
\item[(a)] If $x\nu\nsup_{t_0}(dx)$ converges weakly to a measure
  $x\hat\nu_0(dx) $
with $\int_E x \hat\nu_0(dx)=1$, then for every $t\ge t_0$ we have
that $x\nu\nsup_t(dx)$ converges weakly to $x\nu_t(dx)$, the time-$t$ solution
with initial data $\nu_{t_0}=\hat\nu_0$.
\item[(b)] For any $t\ge t_0$, if $x\nu\nsup_{t}(dx)$ converges weakly to a
measure $x\hat\nu(dx)$ with $\int_Ex\,\hat\nu(dx) =1$, then 
$x\nu\nsup_{t_0}(dx)$ converges weakly to a measure $x\hat\nu_0(dx)$ with
$\int_E x\,\hat\nu_0(dx)=1$, and $\hat\nu=\nu_t$, the time-$t$ solution 
with initial data $\nu_{t_0} = \hat\nu_0$.
\end{enumerate}
\end{thm}
\begin{proof}
We prove (a); the proof of (b) is similar.
 Let  $\sm\nsup_t(dx)=x^2\nu_t(\rme^tdx)$, and with $s=\rme^t$ let
\begin{equation} \label{A.psin1}
\psi\nsup(s,q) = \int_E y^{-2}(\rme^{-qy}-1+qy) \sm\nsup_t(dy).
\end{equation}
The family $\psi\nsup(s_0,\cdot)$ is uniformly
Lipschitz, since equation \qref{A.psidivg} implies
\begin{equation}\label{A.lip}
\psi\nsup(s_0,0)=0,\quad 
0\leq \D_q\psi\nsup(s_0,q) \le 1/s_0 \quad\mbox{ for all $q>0$.}
\end{equation}
The hypothesis is equivalent to saying that the \smeass\ 
$\sm\nsup_{t_0}$ converge to a non-divergent \smeas\ 
$\hat\sm_0(dx)=x^2\hat\nu_0(dx)$ with $\int_E x\inv\hat\sm_0(dx)=1$.
By Theorem~\ref{LT.Mconv} and the criteria in \qref{R.ginf}-\qref{R.divcon},
this is equivalent to the statement that for all $q>0$,
$\psi\nsup(s_0,q)\to\hat\psi_0(q)$, where $\hat\psi_0$ is the 
(second-order) Laplace exponent for $\hat\sm_0$, with
$\D_q\hat\psi_0(0^+)=0$, $\D_q\hat\psi_0(\infty)=1/s_0$. 
(Note $\D_q\hat\psi$ is the first-order Laplace exponent of $\hat\sm_0$.)
As in \qref{A.imp} we have
\begin{equation}\label{A.imp2}
\psi\nsup(s,q)-\psi\nsup(s_0,q-(s-s_0)\psi\nsup(s,q))=0.
\end{equation}
For fixed $s,q$, the sequence $\psi\nsup(s,q)$ is bounded, and
any subsequential limit $\psi_*$ must satisfy
\begin{equation}\label{A.psist}
\psi_* - \hat\psi_0(q-(s-s_0)\psi_*)=0,
\end{equation}
due to the equicontinuity of the maps 
$\psi\mapsto\psi\nsup(s_0,q-(s-s_0)\psi)$.
But equation \qref{A.psist} has the unique solution $\psi_*=\psi(s,q)$,
where $\psi$ is the solution of \qref{A.psi4} with
$\psi(s_0,q)=\hat\psi_0(q)$, $q>0$. 
Hence the whole sequence $\psi\nsup(s,q)$ converges pointwise to 
$\psi(s,q)$. Moreover, differentiating \qref{A.psist} yields
$\D_q\psi_*(0)=0$, $\D_q\psi_*(\infty)=1/s$, since $s_0=1$.
Then the conclusion of (a) follows from 
Theorem~\ref{LT.Mconv}, \qref{R.ginf} and \qref{R.divcon}.
\end{proof}

\subsection{The scaling attractor and eternal solutions}
\begin{thm}\label{AT.attractor}
A probability measure $\hat F$  is an element of the scaling attractor $\attr$ for
Smoluchowski's equation with additive kernel $K=x+y$ if and only if 
$\hat F(dx) = x\nu_0(dx)$ for some eternal solution $\nudot$.
\end{thm}
\begin{proof}
Suppose $\hat  F(dx) = x\nu_0(dx)$ for some eternal solution
$\nudot$. We show $\hat  F \in \attr$. 
Pick arbitrary sequences $T_n, b_n \to \infty$, and consider
the sequence of rescaled  eternal solutions
\[ \nu_t\nsup(dx) = b_n^{-1} \nu_{t-T_n} \left(b_n^{-1}
dx\right), \quad t \in \R. \]
The corresponding distribution functions satisfy $F\nsup_{T_n}(b_n x) =
\hat F(x)$ for every $x$. 
Thus, $\hat F\in\attr$ by Definition~\ref{R.attractor}.
%% \int_0^{b_nx} xb_n^{-1} \nu_0(b_n^{-1} dx)  =
%% \int_0^x y \nu_0(dy) = \hat\mu(0,x) \]
%% for every $x$. Thus, Definition~\ref{A.attractor} holds.

To prove the converse, suppose $\hat F \in \attr$. We show that
$\hat F =  F_0$ for some eternal solution $\nudot$. Let $\hat \vp$
correspond to $\hat\nu$ as in \qref{A.vpdef},
and  $\nudot\nsup,T_n, b_n$ be as in
Definition~\ref{R.attractor}. Consider the rescaled measures 
\[
\tilde\nu\nsup_t(dx)=b_n\nu\nsup_{t+T_n}(b_n dx).
\]
This rescaling yields a solution that is defined for $t\ge -T_n$.
By assumption,
\[ \tilde F\nsup_0(x) = \int_0^x y\, \tilde\nu\nsup_0 (dy) =
F\nsup_{T_n}(b_nx) \to \hat F(x),\]
at all points of continuity. By
Theorem~\ref{AT.cts}, this implies that for any $N\in\N$ the solutions
$\nu\nsup_t$ converge weakly to $\nu_t$ for all $t \geq -N$. In
particular, $\nu_t$ is a solution for $t \geq -N$ for all $N$, thus it is an
eternal solution. 
\end{proof}

Let us now prove that $\attr$ is invariant (part (a) of
Theorem~\ref{RT.AE}). The proof is substantially the same as for
$K=2$. Suppose $\nu$ is a solution on some time interval $[t_1,\infty)$, 
normalized so $ \int_E x\nu_t(dx)=1, t \geq t_1$. 
Suppose $F_{T}\in\attr$ for some $T\ge t_1$. We may presume $T=0$
without loss (if not, we translate in time,
replacing $\nu_t(dx)$ by $\nu_{t-T}(dx)$). 
By Theorem~\ref{AT.attractor} above, $F_T=x\tilde\nu_0$ for 
some eternal solution $\tilde\nu$. But then $\nu_t=\tilde\nu_t$
for all $t\ge t_1$, meaning that $\nu$ is (the restriction of) an
eternal solution. We obtain that $F_t\in\attr$ for every $t \in \R$ by
a similar argument.

\subsection{\LK\/ representation of eternal solutions}
We now prove Bertoin's \LK\/ representation for eternal solutions. 
The proof mainly follows~\cite{B_eternal}, and 
is included to stress the basic framework. 

\begin{thm}[cf.\ Bertoin~\cite{B_eternal}]
\label{AT.eternal}
\begin{enumerate}
\item[(a)] Let $\nu$ be an eternal solution to Smoluchowski's
equation with $K=x+y$, and let $\sm_t(dx)=x^2\nu_t(\rme^tdx)$ be
associated \smeass. 
Then there is a unique divergent \smeas\  $\dsm$
such that $\sm_{t}$ converges to $\dsm$ as $t\to -\infty$. 
%
%(dx)=\sigma^2 \delta_0(dx) +
%x^2\,\Lambda(dx)$ where  $\delta_0$ is a delta mass at 0 and
%$(\sigma^2,\Lambda)$ is a \Levy\ pair.  
%Equivalently, $\psi(s,q) \to \Psi(q)$ as $s\to 0$ for all $q\ge0$. 
\item[(b)] Conversely, given a  divergent \smeas\  $\dsm$  
there  is a unique eternal solution with the properties in part (a),
  defined as follows. Let
\begin{equation}\label{A.Psidef}
\Psi(q) = \int_{\Ebar} \frac{\rme^{-qx}-1+qx}{x^2} H(dx)
\end{equation}
be the Laplace exponent of $H$,
 and let $\psi=\psi(s,q)$ be the solution to 
\begin{equation}\label{A.Imp}
\psi - \Psi(q-s\psi) = 0.
\end{equation}
%\qref{A.psi3} with $\psi(0,q)=\Psi(q)$. 
Then $\nu_t$ is determined by \qref{A.psi2} and \qref{A.psi3}.
\end{enumerate}
\end{thm}
%% \begin{rem}
%% Observe that when $\sigma=0$, the procedure in (b) formally yields
%% a solution to Smoluchowski's equation 
%% with ``initial'' size distribution $\nu_0(dx)=\Lambda(dx)$ having infinite
%% total mass. [[?? At $-\infty$?]]
%% \end{rem}
\begin{proof}
We first prove (a). By Theorem~\ref{LT.Mconv} and
\qref{R.ginf}-\qref{R.divcon}, it is enough  to
show that $\psilevy(q):=\lim_{s \to 0} \psi(s,q)$ exists for every 
$q \geq 0$, with $\D_q\Psi(0)=0$ and $\D_q\Psi(\infty)=\infty$.  
%$\lim_{q \to \infty} q\inv  \psilevy(q)=\infty$. 
We know $\psi\ge0$ and $\partial_q\psi\ge0$, so
 $\partial_s \psi(s,q) \leq 0$ for all $q$, $s$. 
Hence it suffices to show that for each $q>0$, $\psi(s,q)$ stays bounded
as $s\dnto0$.

{\em 1.\/}  We first show $\psi(s,q)$ stays bounded for $q$ near $0$. 
Choose $q_1 >0$ such that $q_{*}:=q_1 -\psi(1,q_1) =\vp(0,q_1) >0$. 
Then $\psi(s,q)=\psi(1,q_1)$ along the characteristic line joining
$(0,q_{*})$ and $(1,q_1)$, so $0\le\psi(s,q)\le\psi(1,q_1)$ whenever
$0<s<1$ and $0<q\le q_{*}$.
%Since $\psi(1,\cdot)$ is convex, the geometry of characteristics is as
%in Figure~\ref{AF.char}. 
(See Fig.~\ref{AF.char}.)
%-----------------------------------------
\begin{figure}
\la{AF.char}
\centerline{\epsfysize=7cm{\epsffile{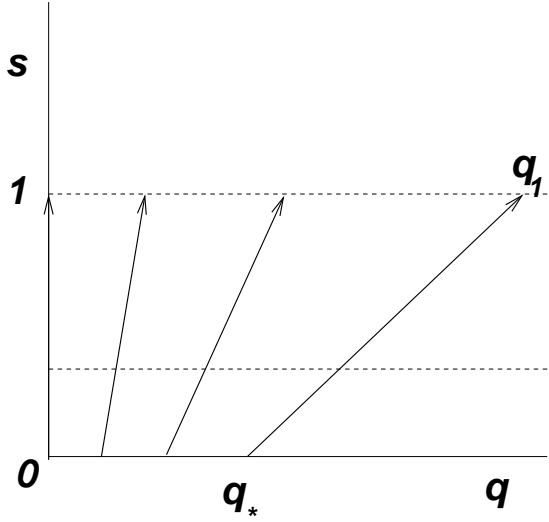}}}
\caption{Geometry of characteristics}
\end{figure}
%------------------------------------------
%Thus, for any $s_0\in(0,1)$, $q_0<??$
%\[ \psi(s_0,q_0) = \psi(1,q(1;s_0,q_0)) \leq \psi(1,q_*), \quad 0 < s_0
%\leq 1.\]
% q_0 \leq q_* -\psi(1,q_*). \]

{\em 2.\/} For $q > q_*$ the complete monotonicity of $q\mapsto q^{-2}
\psi(s,q)$ implies $\psi(s,q) < q^2q_*^{-2} \psi(s,q_*)$. 
%Thus, $ \lim_{s \to 0}  q^{-2} \psi(s,q) = q^{-2}\Psi(q) \leq
%q_*^{-2}\psilevy(q_*)$.

{\em 3.\/} We now show  $\D_q\Psi(0)=0$ and $\D_q\Psi(\infty)=\infty$.
%$\lim_{q \to \infty}q^{-1}\Psi(q)=\infty$. 
Observe that $\Psi$ solves
\[ \psilevy(q)=\psi(1,q+\psilevy(q)), \quad q> 0.\]
Therefore, 
\begin{equation}
\label{A.derivative}
\partial_q \psilevy(q) = \frac{\partial_q\psi(1,q+\psilevy(q))}{1
  -\partial_q\psi(1,q+\psilevy(q))}. 
\end{equation}
Since $\psi(1,q)=q-\vp(0,q)$, we have
$\partial_q \psi(1,q)=1-\partial_q \vp(0,q) \to 0$
as $q\to0$, $\to 1$ as $q \to \infty$. Thus, 
$\D_q\Psi(0)=0$ and $\D_q\Psi(\infty)=\infty$.
%$\lim_{q \to \infty} \partial_q \psilevy(q) = \infty$.
%This also implies $\lim_{q \to \infty} q\inv
%\psilevy(q)=\infty$. Indeed, for any $n \in \N$ choose $q_n$ such that 
%$\partial_q \psilevy(q) \geq n$ for $q \geq q_n$. Then,
%\[\psilevy(q) \geq \psilevy(q_n) + n (q -q_n), \quad q \geq q_n\]
%and $\liminf_{q \to \infty} q\inv \psilevy(q) \geq n$. 
% Since $n$ is
% arbitrary, this proves $ \lim_{q \to \infty} q\inv
% \psilevy(q)=\infty.$.
This proves (a).

We now prove (b). Let $\dsm$ be a divergent \smeas\  and $\Psi$
be defined by \qref{A.Psidef}. Note $\D_q\Psi(0)=0$ and
$\D_q\Psi(\infty)=\infty$ by \qref{R.ginf}-\qref{R.divcon}.
Since $\D_q\Psi(q)>0$, $\psi(s,q)$ is globally defined and
analytic with $\psi(s,q)<q/s$, and \qref{A.psi4} holds for all
$s>0$, $q>0$. With $\Vp$ and $\vp$ defined by \qref{A.psi1},
\qref{A.1} follows. 

By the well-posedness theory in \cite{MP1}, 
we obtain an eternal solution through \qref{A.Fvp},
provided we show that $\D_q\vp(t,\cdot)$ is completely monotone,
which implies that it is the Laplace transform of a (positive) measure
that we call $x\nu_t(dx)$.
From \qref{A.Imp} we obtain that $\vp=\vp(t,q)$ satisfies
\be
\la{A.psi11}
q = \vp + \psilevy(s\vp),
\ee 
whence
\be
\la{A.psi12}
\D_q \vp = \frac{1}{1+s\psilevy'(s\vp)}.
\ee
Since $q\mapsto 1+s\psilevy'(sq)$ is positive with completely monotone
derivative, the map $q\mapsto (1+s\psilevy'(sq))^{-1}$ is
completely monotone \cite[XIII.4]{Feller}. We then infer that 
$\D_q\vp(s,\cdot)$ is completely monotone by Lemma~\ref{AL.comp} below.
Since $\Psi'(0)=0$ we have the normalization \qref{A.mass1}, $\D_q\vp(t,0)=1$.
This finishes the proof of existence. 

Note that total number of clusters $\nu_t(E)=\vp(t,\infty)=\infty$
always here.

Let us show that the eternal solution defined by
this procedure is unique. Let $H$ be a divergent \smeas\  and suppose
$\nu, \tilde{\nu}$ are two eternal solutions with \smeass\  $G_t,
\tilde{G}_t$ that converge to $H$. But this is equivalent to
pointwise convergence of $\psi(s,q)$ and
$\tilde{\psi}(s,q)$ to $\Psi(q)$ as $s\to 0$ where $\psi$ and
$\tilde{\psi}$ solve \qref{A.psi4}. But the solutions to the inviscid
Burgers equation with increasing initial data are unique, thus
$\psi(s,q)=\tilde{\psi}(s,q)$ and $\nu = \tilde{\nu}$.
\end{proof}

\begin{lemma}\la{AL.comp}
Suppose $f,g\colon E\to E$, $f'=g(f)$ and $g$ is completely
monotone. Then $f'$ is completely monotone.
\end{lemma}
\begin{proof}
We prove by induction that the first $n$ derivatives of
$G\circ f$ alternate in sign for every
completely monotone function $G$. For $n=0$, $G(f)>0$. Suppose
the statement is true for some $n\ge0$. Let $G$ be completely
monotone, and note
\[
-(G\circ f)' = -G'(f)g(f)= \tilde G(f)
\]
and $\tilde G$ is completely monotone since it is the product of
completely monotone functions. Using the induction hypothesis, 
we deduce that the first $n+1$ derivatives of $G\circ f$ 
alternate in sign. 
\end{proof}

To complete the proof of Theorem~\ref{RT.LK} for $K=x+y$, we need to
check that the map $\nu_0 \mapsto \dsm$ from $\attr$ to $\Mcan$ 
is a bi-continuous bijection. 
\begin{thm}\label{AT.Mconv2}
Let $\nu\nsup$ be a sequence of eternal solutions with corresponding
divergent \smeass\  $\dsm\nsup$. 
Fix $t \in \R$. Then, taking $n\to\infty$, the following are equivalent:
\bit
\item[(i)] 
$x\nu\nsup_t$ converges weakly to $x\hat\nu$ with $\int_E x \hat\nu(dx)=1$. 
\item[(ii)]  $\dsm\nsup$ converges to a divergent 
\smeas\  $\dsm$.
\eit
If either (equivalently both) of these conditions hold, then 
$\hat\nu=\nu_t$ for an eternal solution with \smeas\  $\dsm$.
\end{thm}
\begin{proof}
With Theorem~\ref{LT.Mconv} in hand, the proof of
Theorem~\ref{AT.Mconv2} is essentially the same as that 
of Theorem~\ref{AT.cts}. 
Assume (i), so $\nu\nsup_t$ converges to $\hat\nu$ with
$\int_E x\hat\nu(E)=1$.  Then $\sm\nsup_t(dx)=x^2\nu_t\nsup(\rme^t dx)$
converges to the \smeas\/ $\hat\sm(dx)=x^2 \hat\nu(\rme^t dx)$ and
the associated Laplace exponents converge: $\psi\nsup(s,q)\to\hat\psi(q)$
for all $q>0$, with $\D_q\hat\psi(0)=0$, $\D_q\hat\psi(\infty)=1/s$. 
Recall that $\psi\nsup (s,q)$ solves
\begin{equation} \label{A.Pconv}
\Psi\nsup(q-s\psi\nsup(s,q)) = \psi\nsup(s,q).
\end{equation}
Let $M >0$. A calculation as in \qref{A.derivative} shows that
$\partial_q\Psi\nsup(q)$ is uniformly bounded in $n$ for $q \in
[0,M]$.  We claim that $\lim_{n \to \infty} \Psi\nsup (q - s\hat\psi(q))$ exists
for every $q$. Let us restrict attention to $q \in [0,M]$. Then by \qref{A.Pconv}
\[ \Psi\nsup(q-s\hat\psi(q)) = \psi\nsup(s,q) + \left(
\Psi\nsup(q-s\hat\psi(q))
-\Psi\nsup(q-s\psi\nsup(s,q))\right). \]
The first term converges to $\hat\psi(q)$ and the second to zero by the
uniform estimate on $\partial_q\Psi\nsup(q)$ on $[0,M]$. Since $M >0$
was arbitrary, we may use
Theorem~\ref{LT.Mconv} to deduce that $\Psi\nsup(q)$ converges to a
Laplace exponent $\Psi(q)$ that satisfies 
\[ \Psi(q-s\hat\psi(q))=\hat\psi(q).\] 
As with \qref{A.derivative} and its sequel it follows that 
$\D_q\Psi(0)=0$ and $\D_q\Psi(\infty)=\infty$.
%$\lim_{q\to \infty} \partial_q\Psi(q) = \lim_{q \to \infty} q^{-1}
%\Psi(q)=\infty$. Thus [[not done??]], 
Thus $\Psi$ is the Laplace
exponent of a  divergent \smeas\/ $\dsm$, and $\dsm\nsup$ converges
to $\dsm$.

We now show (ii) implies (i). Suppose the divergent \smeass\  $\dsm\nsup$
converge to a divergent \smeas\  $\dsm$. Then
Theorem~\ref{LT.Mconv} implies $\Psi\nsup(q)\to\Psi(q)$ for every
$q>0$, and  $\D_q\Psi(0)=0$, $\D_q\Psi(\infty)=\infty$.
%$\partial_q\Psi(q)\to\infty$ as $q\to\infty$. 
Then the characteristics emanating from $s=0$ converge because $q + s
\Psi\nsup(q) \to q + s\Psi(q)$. Thus, $\psi\nsup(s,q) \to \psi(s,q)$,
which satisfies \qref{A.Imp}. 
This yields weak convergence of $x\nu\nsup_t$ to
$x \nu_t$, where $\nu$ is the eternal solution with divergent \smeas\  $\dsm$.
\end{proof}

\subsection{Scaling limits and initial tails}

Let us now prove Theorem~\ref{RT.initial_tails} for the additive kernel.

\begin{proof}[Proof of Theorem~\ref{RT.initial_tails}]
%Let $a_n=\rme^{T_n}$, $b_n=\beta_n$.
We rescale solutions via
$\tilde\nu\nsup_t(dx)=\beta_n\nu\nsup_{t+T_n}(\beta_ndx)$, 
and let $\tilde F\nsup_t(dx) = \beta_n x \tilde\nu\nsup_t(dx)$. 
Also let $\tilde\sm\nsup_t(dx)= x^2\tilde\nu\nsup_t(\rme^t dx)$ and let 
$\tilde\psi\nsup(s,q)$ be the associated Laplace exponent as in
\qref{A.psi2}. Observe
$\tilde\sm\nsup$ in \qref{R.initG} is $\tilde\sm\nsup_{-T_n}$ and 
$\int_E x^{-1} \tilde \sm\nsup(dx) = \rme^{T_n}$.
Let $H$ be the divergent \smeas\  corresponding to $\nu$ and
$\Psi$ its Laplace exponent, and let $\psi(q)$ be the Laplace exponent
of $\sm(dx)= x^2\nu_0(dx)$.

Then (i) is equivalent to saying $\tilde F\nsup_0\to \hat F$ weakly,
meaning the \smeass\/ $\tilde\sm\nsup_0$ converge to $\sm$
with $\int_E x\inv\sm(dx)=1$. This is equivalent to saying
\begin{equation}\label{A.ps1}
\tilde\psi\nsup(1,q)\to \psi(q), \quad q>0, \quad 
\mbox{where $\D_q\psi(0^+)=0$, \  $\partial_q\psi(\infty)=1$.} 
\end{equation}
On the other hand, since $\beta_nx^2\nu\nsup_0(\rme^{-T_n}\beta_ndx)= 
\tilde\sm_{-T_n}(dx)$, 
(ii) is equivalent to saying 
\begin{equation}\label{A.ps2}
\tilde\psi\nsup(\rme^{-T_n},q) \to \Psi(q), \quad q>0, \quad
\mbox{where $\D_q\Psi(0^+)=0$, \  $\partial_q \Psi(\infty)=\infty$.}
\end{equation}
For brevity, let $\tilde\psi\nsup (q)$ denote $\tilde\psi\nsup(1,q)$
and $\tilde\Psi\nsup(q)$ denote $\tilde\psi\nsup(\rme^{-T_n},q)$. Then
the implicit solution formulas to \qref{A.psi4} read
\[ \psi(q) = \Psi(q-\psi(q)), \quad \psi\nsup(q) = \Psi \nsup (
q-(1-\rme^{-T_n}) \psi\nsup(q)). \] 
As in the proof of Theorem~\ref{AT.Mconv2} we may now deduce that 
\qref{A.ps1} is equivalent to \qref{A.ps2}, implying (i) is equivalent
to (ii). The details are omitted. 
\end{proof}

\section{The multiplicative kernel}
\label{sec:mult}
In this section we study scaling dynamics approaching the gelation time
for the kernel $K=xy$.  The study of the multiplicative kernel
can be reduced to the additive kernel by a simple change of
variables. This trick is well-known (see~\cite{Drake}), and allows us
to avoid separate proofs.

\subsection{Solution by the Laplace transform}
The self-similar solutions for $K=xy$  have infinite number and mass,
but finite second moment. However, one 
may develop a natural well-posedness theory using only the finiteness
of the second moment~\cite{MP1}. We assume $\nu_{t_0}$ is a (possibly
infinite) measure with $\int_E x^2 \nu_{t_0}(dx) < \infty$.  Without
loss of generality, we may scale so that $\int_E x^2\nu_{t_0}(dx)=1$
and $t_0=-1$ as in \qref{R.normal2}.  We define the Laplace exponent
(note the change from \qref{A.vpdef}) 
\begin{equation}\label{M.vpdef}
\vp(t,q) = \int_E (1-\rme^{-qx}) x\nu_t(dx) , \quad q \geq 0,
%\qquad 
%\vp_1(q) = \int_E (1-\rme^{-qx}) \nu_1(dx).   
\end{equation}
and write $\vp_0(q)=\vp(t_0,q)$. We may substitute \qref{M.vpdef} in the
moment identity \qref{R.moment} to obtain
\begin{equation}
\label{M.1}
\partial_t \vp - \vp \partial_q \vp = 0, \quad t \in (t_0,0).
\end{equation}
Equation \qref{M.1} may be transformed to \qref{A.1} by the following
change of variables. Let $\vp^{\rm add}(\tau,q)$, $\tau \in [0,\infty)$,
denote a solution to \qref{A.1} with initial data $\vp_0(q)$. Then the
solution to \qref{M.1} is given by
\be
\la{M.2}
\vp(t,q) = \rme^{\tau}\vp^{\rm add}(\tau,q), \quad \tau=\log(|t|^{-1}),
\ee
which may also be written in terms of the number density as 
\be
\la{M.3}
x \nu_{t}(dx) = e^\tau \nu^{\rm add}_\tau(dx).
\ee
Conservation of mass \qref{A.mass1} is now replaced by
\be
\la{M.mass1}
\int_E x^2 \nu_t(dx) = |t|^{-1}, \quad t \in [t_0,0),
\ee
and the probability measure $F_t$ associated to $\nu_t$ is defined by
\be
F(t,x) = |t| \int_0^x y^2\nu_t(dy).
\ee
As in \qref{A.psi3} we define the \smeas\/ 
\be
\label{M.psi3}
\sm_t(dx) = x^3 \nu_{t} (|t|^{-1} \,dx) = \sm_\tau^{\rm add}(dx),
\quad t \in [t_0,0),
\ee
and the associated Laplace exponent
\be
\la{M.psi4}
\psi(t,q) = \int_{\Ebar} y^{-2} \left(e^{-qy}-1+qy \right)
\sm_{t}(dy) = \psi^{\rm add}\left(|t|^{-1},q \right).
\ee
%As before, the rescaling is chosen so that $\sm_t(E)=\sm_{t_0}(E)$
%(if initially finite). 
\noindent
The correspondences \qref{M.3}, \qref{M.psi3} and \qref{M.psi4} map
normalized solutions for $K=xy$  on the 
time interval $t \in [-1,0)$ to normalized solutions with $K=x+y$ on the
interval $\tau \in [0,\infty)$. The same change of variables may be
applied to eternal solutions defined on $t \in (-\infty,0)$. 
By consequence, the results established so far for the additive kernel
carry over in an obvious way for the multiplicative kernel.
This yields continuous dependence of solutions on data 
(by Theorem~\ref{AT.cts}),
the correspondence between the scaling attractor and eternal solutions 
(Theorem~\ref{RT.AE}), the \LK\/ representation (Theorem~\ref{RT.LK}),
and how initial tails encode scaling limits (Theorem~\ref{RT.initial_tails}).
For completeness, we make explicit the map
from divergent \smeass\/ to eternal solutions implicit in
Theorem~\ref{RT.LK}(b).   
\begin{thm}
\label{MT.eternal}
Given a  divergent \smeas\/ $\sm$  
there  is a unique eternal solution defined as follows. Let 
\be
\label{M.psi5}
\Psi(q) = \int_{\Ebar} \frac{e^{-qx}-1+qx}{x^2} \dsm(dx),
\ee
and $\psi(t,q)$, $t\in (-\infty,0)$ be the solution to 
\be
\label{M.psi6}
\psi - \Psi (q+ t^{-1}\psi)=0.
\ee
Then $\nu_t$ is determined by \qref{M.psi3} and \qref{M.psi4}.
\end{thm}

\section{Doeblin solutions}
This section is inspired by Feller's treatment of Doeblin's universal
laws and domains of partial attraction \cite[XVII.9]{Feller}.
But apparently we must be content with using more words
to prove fewer results. Our aim is to prove:

%\begin{refthm}[\ref{RT.dense}]
\begin{thm}\label{D.dense}
There exists an eternal solution $\nu$ whose scaling $\omega$-limit
set contains every element of the proper scaling attractor, $\attr$.
\end{thm}
%\end{refthm}
We will show later that $\attrx$ is the closure of
$\attr$ (see Corollary~\ref{EC.dense}). Therefore,
Theorem~\ref{D.dense} establishes Theorem ~\ref{RT.dense}. 

The proof is based on suitably ``packing the tails''
of the corresponding divergent \smeas. The following is
adapted from Feller~\cite[XVII.9]{Feller}. 
Given an \smeas\ $\sm$ and $a,b>0$ we define
a rescaled measure $\sm^{a,b}$ by
%\sm^{a,b}(x)= a\sm(bx)$.
\be
\label{R.linear1}
\sm^{a,b}(x)= a\sm(bx). 
\ee
\begin{lemma}
\label{D.packing}
Let $\sm_k$ be a sequence of \smeass\ with
\be
\label{D.1}
\int_{\Ebar} x^{-1} \sm_k(dx) \leq k. 
\ee
Then there exist sequences $a_k,b_k$ such that $a_k \to 0$, $ a_kb_k
\to \infty$,  
\be
\label{D.2}
\sm := \sum_{k=1}^\infty \sm_k^{a_k^{-1},b_k^{-1}}
\ee
defines an \smeas, and $\sm^{a_k,b_k}-\sm_k$ converges to zero. 
\end{lemma}
The growth assumption \qref{D.1} is included only
for concreteness and implies no real loss of generality. Our main
purpose is to approximate  divergent \smeass.
\begin{lemma}
\label{D.growth}
Let  $\dsm$ be a divergent \smeas.
 There exists a sequence of
\smeass\ $\sm_k$ satisfying \qref{D.1} such
that $\sm_k$ converges to $\dsm$.
\end{lemma}
\begin{proof}[Proof of Theorem~\ref{RT.dense}]
{\em 1.\/} Let $\tilde\nudot\nsup$ be an arbitrary sequence of 
eternal solutions with
corresponding divergent \smeass\ $\tilde{\dsm}\nsup$. 
Partition the integers into infinitely many subsequences,
and choose $\sm_k \to \tilde{\dsm}\nsup$ for $k$ in the
$n$-th subsequence as in Lemma~\ref{D.growth}.

{\em 2.\/} Now define $a_k,b_k$ and $\sm$ as in Lemma~\ref{D.packing}, and 
put \[\dsm = \delta_0 + \sm. \]
$\dsm$ has an atom at the origin, thus is the divergent \smeas\ for an
eternal solution. By construction, $\sm^{a_k,b_k} \to \tilde{\dsm}\nsup$
as $k \to \infty$ along the $n$-th subsequence.  
Moreover, since $a_k \to 0$, under rescaling 
$\delta_0^{a_k,b_k} = a_k\delta_0$ converges to zero. 
Thus, if we take limits
along the $n$-th subsequence,  $\dsm^{a_k,b_k}\to \tilde{\dsm}\nsup$ . 

{\em 3.\/}  We now apply Theorem~\ref{RT.LK} together with \qref{R.linear3}. 
We have $\dsm^{a_k,b_k}=\Hmap(F^{a_k,b_k}_{t_0})$ and
$F^{a_k,b_k}_{t_0}(x)=F_{T_k}(\beta_k x)$ where
\begin{equation}
(T_k,\beta_k) = \left\{
\begin{array}{ll}
(a_kb_k, b_k) &(K=2),\\
(\log(a_kb_k), a_kb_k^2) &(K=x+y),\\
(-(a_kb_k)^{-1}, a_kb_k^2) &(K=xy).
\end{array}\right.
\end{equation} 
%Set $T_k = a_kb_k,
%\log(a_kb_k)$ and $(a_kb_k)^{-1}$ for $K=2,x+y$ and $xy$ respectively. 
%Similarly let $\beta_k = b_k$ for $K=2$ and $\beta_k=a_kb_k^2$ for
%$K=x+y$ and $xy$. 
Observe that $T_k \to \Tmin$ and $\beta_k \to \infty$.  
We take limits along the $n$-th subsequence to 
obtain $F_{T_k}(\beta_k x) \to \tilde{F}\nsup(x)$ at every point of
continuity.  Hence,  {\em given any sequence of eternal solutions $\nudot\nsup$
there exists an eternal solution $\nudot$ whose scaling $\omega$-limit
set contains each $\tilde{\nu}\nsup_1$}.

{\em 4.\/} The space of divergent \smeass\ is separable. The
\smeass\ which are concentrated at finitely many 
rational points (including 0) with rational weights form a 
countable set which is dense with respect to convergence of \smeass.
By ordering these in a sequence $\tilde{\dsm}\nsup$ and using the
construction above, we see that there exist eternal solutions $\nudot$
such that for {\em every} eternal solution $\tilde\nudot$, $\tilde\nu_1$ is in
the scaling $\omega$-limit set of $\nu$. This finishes the proof of
the theorem.
\end{proof}

\subsection{The packing lemma}
We will need to choose a sequence $c_k$ that grows so fast that
\be
\nn
%\label{D.4}
c_k \sum_{j=k+1}^\infty j c_j^{-1} \to 0.
\ee
The choice $c_k = \rme^{k^2}$ will do. For $j \geq 2$ we have the elementary
estimate
\[ j \rme^{-j^2} < \int_{j-1/2}^{j+1/2} y \rme^{-y^2} dy = \rme^{-j^2-1/4} \cosh j  .
\]
Therefore for $k\geq 1$,  
\[ \rme^{k^2} \sum_{j=k+1}^\infty j \rme^{-j^2} <  \rme^{k^2} \int_{k+1/2}^\infty y
\rme^{-y^2} dy = \frac{\rme^{-k-1/4}}{2} \to 0. \]
\begin{proof}[Proof of Lemma~\ref{D.packing}]
{\em 1.\/} Fix  $a_k b_k = c_k$. Then $\sm$
defines an \smeas\ since
\[ \int_{\Ebar} x^{-1} \sm(dx) = \sum_{k=1}^\infty c_k^{-1}
\int_{\Ebar} x^{-1} \sm_k(dx) \leq \sum_{k=1}^\infty k c_k^{-1} < \infty.\]

{\em 2.\/} Let $\Vp\ksup$ and $\Vp$ denote the Laplace exponents of $\sm_k$ and
$\sm$ respectively. We use the definition \qref{R.linear1} and
\qref{D.2} to obtain 
$\Vp(q) = \sum_{j=1}^\infty c_j^{-1} \Vp\jsup (qb_j)$. Observe 
that $\sm^{a_k,b_k} - \sm_k$ is a positive measure with Laplace
exponent
\[ c_k \Vp(qb_k^{-1}) - \Vp\ksup(q)= c_k\sum_{j \neq k} c_j^{-1}
\Vp\jsup(q b_jb_k^{-1}). \]

{\em 3.\/} To prove convergence to zero, it suffices to show that the
right hand side converges to zero for every $q>0$.  We first control
the tail. Since $\int_{\Ebar} x^{-1} \sm_j(dx) \leq j$,  
\[ c_k\sum_{j=k+1}^\infty  c_j^{-1}\Vp\jsup\left(q b_j b_k^{-1}\right) 
\leq c_k\sum_{j=k+1}^\infty j c_j^{-1} \to 0.\]

{\em 4.\/} We now choose $b_k$ inductively to control the first $k-1$ terms in
the range $0 \leq q \leq k$. Suppose $b_1,\ldots, b_{k-1}$
have been chosen. Since $\Vp\jsup(q) \to 0$ as $q\to 0$, we choose
$b_k$ so large  that
\[ a_k =c_k b_k^{-1} \leq \frac{1}{k}, \quad  c_k\sum_{j=1}^{k-1}
c_j^{-1}\Vp\jsup\left(k b_jb_k^{-1}\right) \leq \frac{1}{k}.\]
\end{proof}
\subsection{Proof of Lemma~\ref{D.growth}}
First, suppose $\dsm$ has no atom at the origin. 
Since $\int_{x}^\infty y^{-1} \dsm(dy)  \to 0$ as $x \to \infty$, we 
may choose a decreasing sequence
$\veps_k$ such that $\int_{\veps_k}^\infty y^{-1} \dsm(dy) \leq k$. Let
$\sm_k(dy) = \mathbf{1}_{y > \eps_k} \dsm(dy)$. Clearly, $\sm_k$
satisfies both conditions of Definition~\ref{R.candef}.  

Next, let $\dsm =\delta_0$. In this
case we choose $\sm_k(dx) = (x \log k )^{-1} \mathbf{1}_{x
  \geq k^{-1}}dx$. Then $\sm_k$ satisfies \qref{D.1} as
\[ \int_{\Ebar} x^{-1} \sm_k(dx) =(\log k)^{-1} \int_{k^{-1}}^\infty x^{-2} dx
= k(\log k)^{-1} \leq k, \]
and 
\[ \Vp\ksup(q) = q (\log k)^{-1} \int_{qk^{-1}}^\infty \frac{1 -\rme^{-x}}{x^2}
dx \to q = \Vp(q).\]
The general case follows by superposition of these two special cases.

\section{Scaling-periodic solutions}
%[Scaling fixed points?]
In this section we characterize scaling-periodic solutions 
and show that they are dense in the scaling attractor.
That is, we prove Theorems~\ref{RT.periodic} and \ref{RT.Pdense}.
\subsection{Characterization}
\begin{proof}[Proof of Theorem~\ref{RT.periodic}]
{\em 1.\/} 
Given a scaling-periodic solution, a solution satisfying \qref{R.periodic}, 
we can scale it as in \qref{newnu}--\qref{newF} so that $t_0$ is as in
\qref{R.normal3}.  Then, under the map
$F \mapsto F^{a,b}$ %from \qref{R.linear3}, 
given by 
\begin{equation}
\label{R.linear3}
F^{a,b}_t(x) =\left\{ \begin{array}{ll}
F_{abt}(bx) & (K=2),\\ 
F_{t+\log(ab)}( ab^2x) & (K=x+y),\\ 
F_{t/ab}(ab^2x) & (K=xy), 
\end{array} \right.
\end{equation} 
for some $a,b>0$ we have $F_t = F^{a,b}_t$ for all $t\in[t_0,\Tmax)$.
Explicitly,
\begin{equation}
(ab,b)= \left\{
\begin{array}{ll}
(t_1,\betarsc) &(K=2),\\
(\rme^{t_1},\betarsc \rme^{-t_1}) &(K=x+y),\\
(-t_1^{-1}, -\betarsc t_1) &(K=xy).
\end{array}\right.
\end{equation}
%A scaling periodic solution must be eternal for
%\qref{R.periodic} may be iterated infinitely many times. 
%If an eternal solution satisfies \qref{R.periodic} it must be a fixed
%point of the map 
%$F \mapsto F^{a,b}$ from \qref{R.linear3} 
%Explicitly, we have $ab= t_1, \rme^{t_1}$, or
%$-t_1^{-1}$; $b=\betarsc,\betarsc \rme^{-t_1}$, or $-\betarsc t_1$; for
%$K=2,x+y$ and $xy$ respectively. 
Observe that $ab>1$ in all three cases. 
Iterating the map, we get that the solution must be (the restriction of)
an eternal solution.
By Theorem~\ref{RT.scale} and \qref{R.linear3}, $F=F^{a,b}$ is equivalent to
$\dsm=\dsm^{a,b}$, that is, 
\be
\label{P.1}
\dsm(x) = a\dsm(bx), \quad x >0.
\ee
Without loss of generality we may suppose $b > 1$ since \qref{P.1} is
equivalent to $a^{-1}\dsm(b^{-1} x) = \dsm(x)$. 

{\em 2.\/} Equation  \qref{P.1} implies $\dsm(0_+)=a\dsm(0_+)$. If $\dsm$ has
an atom at the origin, this forces $a=1$. Then $\dsm(x)=\dsm(bx)$ for every
$x>0$, and since $b>1$ and $\dsm(x)$ is non-decreasing, 
it follows $\dsm(x)=c=\dsm(0_+)$ for all $x>0$. 
Therefore, if $\dsm$ has an atom at the origin, then $\dsm=c \delta_0$
for some $c>0$.

{\em 3.\/} Suppose $\dsm$ does not have an atom at the origin. 
We iterate \qref{P.1} to find that 
\[
\int_1^{b_-}\dsm(dx)=a^j\int_{b^j}^{b_-^{j+1}}\dsm (dx), \quad
\int_1^{b_-}\frac{\dsm(dx)}{x} = (ab)^j\int_{b^j}^{b_-^{j+1}}\frac{\dsm (dx)}{x}.
\]
In order that $\dsm$ is an \smeas\ we require
\[ \int_{E} (1\wedge x^{-1})\dsm(dx) = \sum_{j<0} a^{-j}
\int_1^{b_-} \dsm(dx) + \sum_{j \geq 0} (ab)^{-j}
\int_1^{b_-}x^{-1}\dsm(dx) <\infty.
\]
Thus, $a <1$ and $ab>1$.  Given $x>0$ let $k= \max\{j:  b^j \leq
x\}$. A similar calculation yields
\[
\dsm(x) =   \sum_{j < k}a^{-j} \int_1^{b_-}\dsm(dx)  + a^{-k} \int_1^{b^{-k}x_-} \dsm(dx).  \]
This shows $\dsm$ is determined by its restriction to $[1,b)$. 

{\em 4.\/} Conversely, suppose $\dsm=\dsm^{a,b}$ and (i) or (ii) hold. 
Notice that $\dsm$ is automatically divergent since it either has an
atom at the origin or 
\[ \int_E x^{-1} \dsm(dx) = \int_1^{b_-} x^{-1}\dsm(dx) \sum_{j
  =-\infty}^\infty (ab)^{-j}= \infty.\]
Thus, it determines an eternal solution, which by
\qref{R.linear3} satisfies $F =F^{a,b}$.
\end{proof}

\subsection{Self-similar solutions}
As remarked in Section~\ref{RS.chaos}, the case (i) 
is simple but important. The associated divergent \smeas\ is
scale-invariant for every $b>1$ and the scaling-periodic solutions
are the classical self-similar solutions with exponential tails.
%\begin{equation}\label{C.sss1}
%\nu_t(dx)=c^{-1}t^{-2}\rme^{-x/c t}\,dx.
%\end{equation}
If a scaling-periodic solution satisfies \qref{R.periodic} for
every $t_1 >t_0$ (with changing $\betarsc$), it follows that 
for some fixed $a$ and $b$, $\dsm(x)=a^{r}\dsm(b^rx)$ for all rational and
hence all real $r$. The fundamental rigidity lemma for scaling
limits~\cite[VIII.8]{Feller} then implies $\dsm(x)= C_\theta x^\theta$
for some $\theta \in \R$. The finiteness condition $\int_E (1 \wedge
x^{-1}) \dsm (dx) < \infty$  then implies $\theta =1-\rho, \rho \in
(0,1]$. If $\rho=1$, $\dsm$ is an atom at the origin corresponding to
(i) above. The self-similar profiles and their domains of attraction
are discussed further in Section~\ref{sec:domains}.

\subsection{Density of scaling-periodic solutions}
% 
% Let $\nu$ be an eternal solution to Smoluchowski's equation with
% kernel $K=2$, $x+y$ or $xy$.
% Let $a_n \to 0,b_n \to \infty $ be sequences such that $a_nb_n^{1/2}
% \to 0$ and $a_nb_n\to\infty$. 
% Then there exist scaling-periodic solutions $\nu\nsup$ with scale parameters
% $(a_n,b_n)$ such that $\nu\nsup_t$ converges weakly to $\nu_t$ for
% every $t \in (\Tmin,\Tmax)$.
To prove Theorem~\ref{RT.Pdense} and establish density of 
scaling-periodic solutions in the full scaling attractor $\attrx$, 
it will suffice to prove such solutions are dense in the proper
scaling attractor $\attr$ (see Corollary~\ref{EC.dense}).
%\begin{refthm}[\ref{RT.Pdense}]\label{CT.Pdense} 
%Let $\hat F\in\attrx$
%be arbitrary.  Let $a_n \downarrow 0,b_n \uparrow \infty $ be
%sequences such that $a_nb_n^{1/2} \to 0$ and $a_nb_n\to\infty$.  Then
%there exist scaling-periodic solutions $\nu\nsup$ with scale
%parameters $(a_n,b_n)$ such that $x^\gamma\nu\nsup_{t_0}(dx)$
%converges weakly to $\hat F$ as $n\to\infty$.  
%\end{refthm}
\begin{thm}
\label{PT.Pdense}
Scaling-periodic solutions are dense in $\attr$.
\end{thm}
\begin{proof}%[Proof of Theorem~\ref{RT.Pdense}]
{\em 1.\/} Let $\hat F\in\attrx$ be arbitrary.  Let $a_n \downarrow
0,b_n \uparrow \infty $ be sequences such that $a_nb_n^{1/2} \to 0$
and $a_nb_n\to\infty$.  We claim that there exist scaling-periodic
solutions $\nu\nsup$ with scale parameters $(a_n,b_n)$ such that
$F\nsup_{t_0}(dx)=x^\gamma\nu\nsup_{t_0}(dx)$ converges weakly to $\hat F$ as
$n\to\infty$.  
Let $\dsm$ denote the divergent \smeas\ associated with $\nu$.
By Theorems~\ref{RT.LK} and \ref{RT.scale} it suffices to construct
divergent \smeass\  $\dsm\nsup$ such that 
$\dsm\nsup=a_n \dsm\nsup(b_n\,\cdot)$ and $\dsm\nsup$ 
converges to $\dsm$.

{\em 2.\/}  Consider first the case where $\dsm$ has no
atom at the origin. In this case we define $\dsm\nsup$ to be the
scaling-invariant extension 
of $\dsm$ restricted to the interval $I_n:=[b_n^{-1/2},b_n^{1/2})$.
Then for any $x>0$ that is a point of continuity of $\dsm$, for 
$n$ large we have $x\in I_n$ and  
\[
\dsm\nsup(x)= \int_{b_n\inv}^{x}  \dsm(dx) + \sum_{j<0} (a_n)^{-j}
\int_{b_n\inv}^{1} \dsm(dx)
\to \dsm(x)
\]
as $n\to\infty$. Moreover,
\[
\int_x^\infty \frac{\dsm\nsup(dy)}{y} = 
\int_x^{b_n} \frac{\dsm(dy)}{y} + 
\sum_{j\ge1} (a_nb_n)^{-j} \int_1^{b_n}\frac{\dsm(dy)}{y}
\to \int_x^\infty \frac{\dsm(dy)}{y}.
\]
This establishes the desired convergence of \smeass\/.

{\em 3.\/} In case $\dsm=\delta_0$, we let $\dsm\nsup$ be a sum
of delta masses $\delta\nsup_j, j \in \Z$ concentrated at points 
$\beta_j = b_n^{j-1/2}$, so
that $\dsm\nsup = \sum_{j} (a_n b_n)^j \delta\nsup_j$.
Observe that there is no mass in $(b_n^{-1/2}, b_n^{1/2})$; 
thus for any $x>0$, for $n$ large we have
\[
\dsm\nsup(x) = \sum_{j\le0} (a_nb_n)^j = \frac{1}{1-a_nb_n} \to 1,
\]
and 
\[
\int_x^\infty y\inv \dsm\nsup(dy) = b_n^{1/2} \sum_{j>0} a_n^j =
\frac{a_n b_n^{1/2}}{1-a_n} \to 0. 
\]
Hence the \smeass\/ $\dsm\nsup$ converge to $\delta_0$.

{\em 4.\/} In the general case, we simply superpose the separate constructions. 
Observe that the restriction $a_n b_n^{1/2}\to0$ is only needed in the
critical case when $\dsm$ has an atom at the origin.
\end{proof}

\section{Extended solutions, with dust and gel}
\label{S.extend}
\subsection{Extended solutions}
A proper solution to
Smoluchowski's equation satisfies $\int_E x^\gamma\nu_t(dx)=
m_\gamma(t)$ with $m_\gamma(t)$ normalized as in \qref{R.mgamma}. 
However, a sequence of proper solutions may lose mass in the
limit. We append atoms at $0$ and $\infty$ to account for these
defects, considering measures on $\Ebar=[0,\infty]$ of the form
\be
\label{E.ext1}
\mu_t = \dust(t)\delta_0 + x^\gamma \nu_t + \gel(t)
\delta_\infty, 
\ee
where $\nu_t$ is a size-distribution measure on $E$. We call the atoms $\dust$
and $\gel$ the dust and gel respectively. An associated probability
measure on $\Ebar$ is defined as in \qref{R.Fp}, by
\be
\label{E.ext2}
\Fbar_t(dx)= \frac{\mu_t(dx)}{\mu_t(\Ebar)} 
= \frac{\dust(t)\delta_0(dx) + x^\gamma \nu_t(dx) + \gel(t) \delta_\infty(dx)}
{\dust(t)+\int_E x^\gamma\nu_t(dx)+ \gel(t)} .
%\int_{[0,x]} y^\gamma \nubar_t(dy) \left\slash
%\int_{\Ebar} y^\gamma\nubar_t(dy). \right.
\ee
The \smeas\/ associated to a solution in \qref{R.cansoln1} is 
replaced by the \barsmeas
\be
\label{E.ext3}
\esm_t(dx) = x^{\gamma+1} \nu_t(\lambda(t) \,dx) + \ggel(t)
\delta_\infty(dx), \qquad \ggel(t) = \frac{\gel(t)}{\lambda(t)^\gamma}. 
\ee
The measures $\mu_t$ define Laplace
exponents by evident modification of equation 
\qref{C.vpdef} for $K=2$, namely
\be
\vp(t,q)=\int_{\Ebar}(1-\rme^{-qx})\mu_t(dx)
= a_\infty(t)+\int_E(1-\rme^{-qx})\nu_t(dx),
%\quad(K=2),
\ee
and of \qref{A.vpdef}  and
\qref{A.psi2} for $K=x+y$ and \qref{M.vpdef} for $K=xy$, both yielding
\be
\vp(t,q)=\int_{\Ebar}\frac{1-\rme^{-qx}}x\mu_t(dx)
= a_0(t)q+\int_E(1-\rme^{-qx})\nu_t(dx).
%\quad(K=x+y)
\ee
%\marginpar{Formulas??}
The evolution
equations for these exponents remain 
\be
\label{E.ext4}
\partial_t \vp = \left\{ 
\begin{array}{rl} -\vp^2, &  (K=2),\\
    \vp \partial_q \vp -\vp,  & (K=x+y),
\\ \vp \partial_q \vp, &  (K=xy).   
\end{array} \right.
\ee
This motivates the following definition.
\begin{defn}
\label{ED.ext}
%Given a probability measure $\specialF$ on $\Ebar$, 
A family of triples $(\nu_t,a_0(t),a_\infty(t))$, $t \in [t_0,\Tmax)$,
defines an {\em extended solution} to Smoluchowski's equation 
for the kernels $K=2$, $x+y$ and $xy$ with initial data 
$(\hat\nu,\hat a_0,\hat a_\infty)$,  if  
\begin{enumerate}
\item[(a)] The measures $\mu_t$ in \qref{E.ext1} satisfy $\mu_t(\Ebar)=
m_\gamma(t)$ with $m_\gamma(t)$ as in \qref{R.mgamma}, 
for $t \in [t_0,\Tmax)$.
%$x^\gamma \nubar_t (\cdot)= m_\gamma(t)\Fbar_t(\cdot)$ 
%with $x^\gamma\nubar_t(\Ebar)= m_\gamma(t)$, 
%$t \in [t_0,\Tmax)$ with $m_\gamma(t)$ as in \qref{R.mgamma}.
\item[(b)] \qref{E.ext4}  holds for $q>0$ and $t \in (t_0,\Tmax)$. 
%with  $x^\gamma\nubar_t=m_\gamma(t)\Fbar_t$.
\item[(c)] $\mu_t \to \hat\mu
=\hat a_0\delta_0+x^\gamma\hat\nu+\hat a_\infty\delta_\infty$ 
weakly as $t \dnto t_0$. 
 \end{enumerate}
\end{defn}
%A family of probability measures $\Fbar_t$, 
Due to the normalization in (a), we regard extended solutions 
as determined by the associated probability distributions 
$\Fbar$ in \qref{E.ext2}.
We will usually denote an extended solution with values 
$(\nu_t,a_0(t),a_\infty(t))$ simply by $\nu$. 

Extended solutions provide  the correct compactification in light of  
the following theorem. Since every proper solution also defines an
extended solution, the
theorem applies in particular to sequences of proper solutions. 
\begin{thm}
\label{ET.const-compact}
Let $\extsol_t\nsup$, $t\in[t_0,\Tmax)$, be probability measures 
associated with a sequence of extended solutions $\nu\nsup$.
Then there exists a sequence $n_j\to\infty$ and 
probability measures $\extsol_t$ associated 
with an extended solution $\nu$, such that 
$\extsol\nsupj_t$ converges weakly to $\extsol_t$
for every $t \in [t_0,\Tmax)$. 
\end{thm}
\begin{proof}
Consider the sequence of probability measures $\extsol\nsup_{t_0}$ on $\Ebar$. 
Then there exists a 
subsequence $n_j$ and a probability measure $\specialF_0$ such that
$\extsol_{t_0}\nsupj$ converges weakly to $\specialF_0$. 
%and $\extsol_{t_0}(\Ebar)=1$. 
We use $\specialF_0$ to determine initial data 
to define an extended solution $\nu$ for $t \in
[t_0,\Tmax)$. Continuous dependence on initial data as in
Theorem~\ref{ET.cts} below implies the weak convergence of
$\extsol\nsupj_t$ to $\extsol_t$ for every $t \in [t_0,\Tmax)$.
\end{proof}

We state the following result without proof, as it is an easy consequence of
Definition~\ref{ED.ext}, and Theorems~\ref{CT.cts}
and~\ref{AT.cts}. The notion of extended 
solution allows us to simplify matters, as it is no longer necessary to 
assume that $\hat\mu_0(\Ebar)=m_\gamma(t_0)$ , or
$\hat\mu(\Ebar) =m_\gamma(t)$ as in parts (a) and (b) of
Theorem~\ref{CT.cts} and ~\ref{AT.cts}. 
\begin{thm}\label{ET.cts}
(Continuous dependence on data.)
For Smoluchowski's equation with kernels $K=2$, $x+y$ or $xy$, let
$t_0 \in (\Tmin,\Tmax)$ and let $\extsol\nsup$ determine a sequence of
extended solutions defined for $t\in I=[t_0,\Tmax)$. 
\begin{enumerate}
\item[(a)] If $\extsol\nsup_{t_0}$ converges weakly to a
 measure $\specialF_0$, then for every $t\in I$,
  $\extsol\nsup_t$ converges weakly to $\extsol_t$, 
associated with the time-$t$  
extended solution with initial data determined by $\extsol_{t_0}=\specialF_0$.
\item[(b)] For any $t\in I$, if $\extsol\nsup_{t}$ converges
  weakly  to a measure $\specialF$, then  
$\extsol\nsup_{t_0}$ converges weakly to a probability measure $\specialF_0$
%$F_0$ with $\extsol_0(\Ebar)=1$, 
and $\specialF=\extsol_t$, associated with the
time-$t$ solution  with initial data determined by 
$\extsol_{t_0} =\specialF_0$. 
\end{enumerate}
\end{thm}

\subsection{Transformation to proper solutions}
Clusters of ``zero'' or ``infinite'' size interact with other clusters
in simple ways.
The invariances of the evolution equations \qref{E.ext4} allow us to
relate all extended solutions (except pure dust and gel) 
to proper solutions. Let us consider the constant and additive kernels in turn.
\subsubsection{The constant kernel}
The dust and gel are recovered as limits as $q \to 0$ and $\infty$
respectively: 
\be 
\label{E.c_gel}
\gel(t) = \vp(t,0^+), \quad \dust(t) = \mu_t(\Ebar)- \vp(t,\infty^-).
\ee
Since $\mu_t(\Ebar)=t\inv$, 
we take limits in \qref{E.ext4} to see that the dust and gel satisfy 
\be
\label{E.c_gel2}
\frac{d\gel}{dt} = -\gel^2, \quad \frac{d{\left(t^{-1}-
      \dust\right)}}{dt}  = -
(t^{-1}-\dust )^2. 
\ee
The extended solution corresponds to purely dust and gel when
$\mu_t(E)=0$, so that $\dust(t)+\gel(t)=t\inv$.
We may exploit \qref{E.ext4}  to show that every extended solution
that is not purely dust and gel is in
correspondence with a proper solution after a simple change of scale. Suppose
$\vp(t,q)$ is the Laplace exponent of an extended solution. 
If $\gel(t_0)>0$ let
%\marginpar{include calculations?}
\be \la{e:ext_rsc1}
\hvp(\htt,q) = \alpha(t)^{-2} \left(\vp(t,q)-\gel(t)\right),
\ee
where
\be
\htt\inv = \alpha(t)^{-2}(t\inv-\dust(t)-\gel(t)), 
\qquad
\alpha(t) = \frac{\gel(t)}{\gel(t_0)}.
\ee
Then we find $\hvp(\htt,0^+)=0$, $\hvp(\htt,\infty^-)=\htt\inv$,
and $\partial_{\htt}\hvp= -\hvp^2$.
Thus, $\hvp(\htt,q)$ is the Laplace exponent of a proper solution 
defined on $[\htt_0,\infty)$. 

For vanishing gel ($\gel(t_0)\to0$) the transformation above
simplifies, yielding $\alpha=1$,  $\htt-\htt_0=t-t_0$, $\hvp=\vp$. 
Zero-size clusters combine trivially with other clusters, 
so the presence of dust only shifts time in accord with our
normalization of total number.
Observe that if gel is present ($\gel(t_0)>0$),
the probability of being gel approaches one ($\gel(t)/\mu_t(\Ebar)\to1$) 
and the relative distribution of finite-size clusters approaches
a state reached by the proper solution at a finite time;
we have $\htt\to \htt_0+ 1/\gel(t_0)$ as $t \to \infty$.

%\ba
%\label{eq:ext_rsc1}
%&& \hvp(\htt,q) = \alpha(t)^{-2}\left(\vp(t,q)- \gel(t) \right), 
%\quad
%\alpha(t) = \frac{\gel(t)}{\gel(t_0)}, \\
%\nn %\label{eq:ext_rsc2}
%&& \htt -\htt_0 = \alpha(t) \left(t-t_0 \right), \quad  \htt_0=
%\frac{t_0}{1 - t_0\left(\dust(t_0)+\gel(t_0)\right)}. 
%\ea
%Then we have $\partial_{\htt}\hvp= -\hvp^2$ 
%and $\hvp(\infty^-,\htt)= \htt^{-1}$. 

\subsubsection{The additive kernel}
In this case, $\D_q\vp(t,q)=\int_{\Ebar}\rme^{-qx}\mu_t(dx)$ and 
$\mu_t(\Ebar)=1$, so the dust and gel are given by
\ba
\label{eq:ext_add2}
\dust (t) = \partial_q \vp(t,\infty), \quad  \gel (t) = 1- \partial_q
\vp(t,0^+). 
\ea
The similarity with the constant kernel is clear if we
use the time scale $s=e^t$ and the Laplace
exponent $\psi(s,q)$ defined in \qref{A.psi1} and \qref{A.psi2}. 
Let
\be
\label{E.ab1}
\bdust(s) = \frac{1}{s} - \partial_q\psi(s,\infty) =\frac{\dust(t)}{s}, \quad
\bgel(s)=\partial_q \psi(s,0) = \frac{\gel(t)}{s}.
\ee
We then take limits in \qref{A.imp} to see that
\be
\label{E.ab2}
\frac{d(s^{-1}-\bdust)}{ds} = -(s^{-1}-\bdust)^2,
\quad 
\frac{d\bgel}{ds} = -\bgel^2, 
\ee
which is equivalent to the following closed
equations for the dust and gel:
\ba
\label{eq:ext_add4} 
\frac{d\dust}{dt} = -\dust (1-\dust), \quad \frac{d\gel}{dt} =
\gel(1-\gel). 
\ea
The extended solution is purely dust and gel when $\dust(t)+\gel(t)=1$. 
If it is not, we exploit the invariances of the inviscid Burgers equation
\qref{A.psi4} to reduce extended solutions to proper solutions
by a change of scale. Given initial data $\psi_0$ with $\partial_q
\psi_0(0) = \bgel(s_0) \geq 0$ and $\partial_q \psi_0(\infty) =
s_0^{-1}-\bdust(s_0) > \bgel(s_0)>0$, we define a proper solution via 
the change of variables
\be \la{e:ext_add5}
 \hpsi(\hs,\hq) = \alpha(s)^{-1}\left(\psi(s,q) -\bgel(s) q\right),
\ee
where
\[
\hs\inv = \alpha(s)^{-2}\left(s\inv-\bdust(s)-\bgel(s)\right), \quad
\hq=\alpha(s)q, \quad
\alpha(s)= \frac{\bgel(s)}{\bgel(s_0)}.
\]
This ensures $\partial_{\hq} \hpsi(s,0) = 0$,
$\partial_{\hq} \hpsi(s,\infty)=s^{-1}$, and 
$\partial_{\hs} \hpsi + \hpsi \partial_{\hq} \hpsi=0$ 
for $\hs > \hs_0$.

%\ba
%\label{eq:ext_add5}
%&& \hpsi(\hs,\hq) = \alpha(s)^{-1}\left(\psi(s,q) -\bgel(s) q\right), \quad
%\alpha(s)= \frac{\bgel(s)}{\bgel(s_0)}\\
%\nn %\label{eq:ext_add6}
%&& 
% \hq = \alpha(s) q ,  \quad \hs-\hs_0 = \alpha(s) (s-s_0), \quad \hs_0 =
%\frac{s_0}{1-s_0\left(\bdust(s_0)+\bgel(s_0)\right)}. 
%\ea
%This ensures $\partial_{\hs} \hpsi +
%\hpsi \partial_{\hq} \hpsi=0$, $\hs > \hs_0$, with initial data 
%such that $\hpsi(\hs_0,0)=0$ and
%$\hpsi(\hs_0,\infty)=s_0^{-1}$. It then follows 
%that $\partial_{\hq} \hpsi(s,0) = 0$ and 
%$\partial_{\hq} \hpsi(s,\infty)=s^{-1}$ for every $\hs > \hs_0$. 

\subsection{\LK\/ representation}
% of extended eternal solutions}
%An extended eternal solution is defined as in \ref{R.eternal}.
\begin{defn}
\label{ED.eternal}
An extended solution to Smoluchowski's equation that is defined for
all $t \in (\Tmin,\Tmax)$ is called an {\em eternal extended 
solution}. 
\end{defn}
The following representation theorem is the completion of
Theorem~\ref{RT.LK}. We establish a bijection between the set of
eternal extended solutions and the space $\Mcanx$ consisting of 
all \barsmeass\/ together with a point at infinity. 
The point at infinity corresponds to all
measures such that $\int_{\Ebar}(1\wedge x^{-1})
\dsm(dx)=\infty$. These measures give rise to the (unique)
Laplace exponents $\Phi(q)=\Psi(q)=\infty$, $q >0$. We say a sequence
of \barsmeass\/ converges to the point at infinity if  $\Vp\nsup(q)
\to \infty$, $q>0$ for the associated Laplace exponents.
This special case corresponds to
the eternal extended solution that is pure gel. It is the counterpoint
to the Laplace exponents $\Phi(q)=\Psi(q)=0$, $q>0$ which generate
the eternal extended solution that is pure dust.
%
%It is worth remarking that in
% the theorem below
%involving the \barsmeass\/ defined in \qref{E.ext3}. This completion
%involves 
%We now consider the space of all \sbar
% Since the proof of the \LK\/ formula relies only on the explicit
% solution formulas and the topology of \barsmeass\/ we obtain the
% following completion of 
% \
\begin{thm}
\label{ET.LK}
\begin{enumerate}
\item[(a)]Let $\nu$  be an eternal extended 
solution for Smoluchowski's equation with
$K=2,x+y$ or $xy$. If $\nu$ is not pure gel, there is an \barsmeas\ $\dsm$ 
such that $\esm_t$ converges to $\dsm$ as $t \dnto \Tmin$. If $\nu$
is pure gel, $\esm_t$ converges to the point at infinity in $\Mcanx$.
\item[(b)] Conversely, for every \barsmeas\ $\dsm$, 
there is a unique eternal extended solution $\nu$ 
  such that $\esm_t$ converges to $\dsm$ as $t \dnto \Tmin$.  The
  point at infinity generates the extended solution $\nu$ that is pure gel.
\item[(c)] 
Let $\Hmapx\colon\attrx\to\Mcanx$
map the (full) scaling attractor $\attrx$
to the set $\Mcanx$ of  \barsmeass\
by $\Hmapx(\specialF)=\dsm$, where $\dsm$ is the \barsmeas\ 
associated to the eternal extended solution $\nu$ such that
$\specialF=\Fbar_{t_0}$ with $t_0$ as in
\qref{R.normal2}. Then $\Hmapx$ is a bi-continuous bijection.
Moreover, $\Hmap\colon\attr \to \Mcan$ is the restriction of $\Hmapx$
to $\attr$.
\end{enumerate}
\end{thm}
The map $\Hmapx$ is defined in terms of Laplace exponents by the same
formulas as for proper solutions: \qref{C.eternal} for $K=2$,
\qref{A.imp} for $K=x+y$, and \qref{M.psi6} for $K=xy$. 
Parts (a) and (b) of the theorem are then proven just as in
Theorems~\ref{CT.eternal} and \ref{AT.eternal}. The proof here is
simpler, since we no longer need verify the divergence conditions on
the \barsmeas\/. The proof of part (c) relies on two separate
arguments. It is easy 
to show as in Theorems~\ref{CT.Mconv2}  and \ref{AT.Mconv2} that the
map $H \mapsto \Fbar_{t_0}$ is a bi-continuous bijection. 
However, we must also identify  such $\Fbar_{t_0}$ as belonging
to the attractor. Here the arguments deviate
slightly from those of Section~\ref{sec:const-ker} and
~\ref{sec:add}. We use parts (a) and (b) of Theorem~\ref{ET.LK} in the
proof of part (c) via the following intermediate theorem. 
% Extended eternal solutions  are in correspondence with elements of the
% (full) scaling attractor defined in \ref{R.full}. 
\begin{thm}
\label{ET.AE}
\begin{enumerate}
\item[(a)]
The scaling attractor $\attrx$ is invariant: If $\nu$ is an extended solution
of Smoluchowski's equation,
and $\Fbar_t\in\attrx$ for some $t$, then $\nu$ is eternal and
$\Fbar_t\in\attrx$ for all $t\in(\Tmin,\Tmax)$.
\item[(b)]
A probability measure $\specialF$ on $\Ebar$ belongs to 
$\attrx$ if and only if $\specialF= \Fbar_{t_0}$
for some extended eternal solution $\nu$. 
\end{enumerate}
\end{thm}
\begin{proof}The proof differs from earlier arguments only in the
  first part of  Theorems~\ref{CT.attr} and~\ref{AT.attractor} (the
  assertion that $\Fbar_{t_0} \in
  \attrx$ if $\nu$ is eternal).  In order to prove this, let us 
  suppose $\nu$ is an 
  extended eternal solution with associated \barsmeas\/ $\bar\dsm =
  (\dsm, g_\infty)$ where $\dsm$ is an \smeas\/ and $g_\infty$ is the 
charge of $y\inv H(dy)$ at $\infty$. 
To show that $\specialF :=\Fbar_{t_0}$ is in 
  the scaling attractor, we must find $T_n\upto \Tmax,\beta_n \to \infty$ and 
a sequence of {\em proper\/} solutions such that
$F\nsup_{T_n}(\beta_nx) \to \specialF(x)$ at points of continuity. We
use the \LK\/ formula to find such solutions. We approximate $\bar{H}$ by 
the sequence of  {\em divergent\/} \smeass\/ $\dsm\nsup = n^{-1}
\delta_0 + \dsm + g_\infty n\delta_n$. It follows that for the corresponding
 (proper) eternal solutions $\nu\nsup$, the probability measures 
$\Fbar_t\nsup$ converge to $\Fbar_t$ for every $t> \Tmin$. 
Given any sequence $T_n \upto \Tmax, \beta_n \to \infty$
 we consider a  sequence of rescaled solutions determined as in
\qref{newF}, by
\[ \tilde{F}\nsup_t(x) = \left\{ \begin{array}{rl}
    F\nsup_{t/T_n}(\beta_n^{-1} x), & (K=2), \\
    F\nsup_{t - T_n} (\beta_n^{-1} x), & (K=x+y), \\
    F\nsup_{t/|T_n|}(\beta_n^{-1}(x)), & (K=xy). \end{array} \right. \]
We then have  $\tilde{F}\nsup_{T_n}(\beta_nx) = F\nsup_{t_0}(x)
\to \specialF(x)$ at all points of continuity. 

The converse implication and part (a) are proven exactly as in
Theorems~\ref{CT.attr} and~\ref{AT.attractor} and the sequel.
\end{proof}
This also  proves a property alluded to several times before.
\begin{cor}
\label{EC.dense}
$\attrx$ is the closure of $\attr$.
\end{cor}
\begin{proof}
If $\specialF \in \attrx$ has \barsmeas\/ $\bar\dsm$, we approximate
$\bar\dsm$ by a sequence of divergent \smeass\/ as above.
\end{proof}

\subsection{Scaling limits and initial tails}
We now state the natural extension of Theorem~\ref{RT.initial_tails}
to eternal extended solutions. The proof is almost identical to that of
Theorem~\ref{RT.initial_tails} except that we no longer need verify
divergence of the \smeas\/.
\begin{thm}
\label{ET.code}
Let $\specialF\in\attrx$ 
%so $\hat F(dx)=x^\gamma\nu_{t_0}(dx)$ where $\nu$ is an eternal solution, 
with associated \barsmeas\ $H$.
Let $\nu\nsup$ be any sequence of proper solutions defined for $t\in
[t_0,\Tmax)$,
with associated initial \smeass\ given by
$\sm\nsup(dx)=x^{\gamma+1}\nu\nsup_{t_0}(dx)$. 
Let $T_n \to \Tmax$, $\beta_n\to\infty$. 
Then the following are equivalent: 
\bit
\item[(i)] ${F}\nsup_{T_n}(\beta_n x) \to \specialF(x)$ 
as $n\to\infty$, at every point of continuity.
\item[(ii)] The rescaled initial \smeass\ $\tilde\sm\nsup$ defined by
\qref{R.initG} converge to the \barsmeas\/ $\dsm$ as $n \to \infty$.
% \begin{equation}\la{R.initG}
% \tilde\sm\nsup(x) = \left\{
% \begin{array}{ll}
% \beta_n^{-1}T_n \,\sm\nsup(\beta_nx)
% %= T_n\beta_n^{-1}\int_{[0,\beta_nx]} y\nu\nsup_{t_0}(dy) 
%          &(K=2),\\[6pt]
% \beta_n\inv \rme^{2T_n}\, \sm\nsup(\rme^{-T_n}\beta_nx) 
% %= \beta_n\inv\rme^{2T_n} \int_0^{x\beta_n\exp(-T_n)}y^2\nu\nsup_0(dy) 
% &(K=x+y), \\[6pt]
% \beta_n\inv |T_n|^{-2} \,\sm\nsup(|T_n|\beta_n x)
%  &(K=xy),
% \end{array} \right.
% \end{equation}
% have the property that
% $\tilde\sm\nsup$ converges to $\dsm$ as $n \to \infty$.
\eit
\end{thm}

\section{Initial tails and ultimate scaling dynamics}
In this section, we present two applications of the principle that
ultimate scaling dynamics are encoded in the initial tails (as
formalized in theorems~\ref{RT.initial_tails} and~\ref{ET.code}). 
The first is a proof of the shadowing theorem~\ref{RT.shad}. The second is a
streamlined proof of the classification of domains of
attraction in~\cite{MP1} that avoids the use of Karamata's Tauberian theorem.  

\subsection{Initial tails and shadowing}
\begin{proof}[Proof of Theorem~\ref{RT.shad}]
{\em 1.\/} As in section~2, we let $\dist(\cdot,\cdot)$ denote any metric on
$\bar\PP$ which induces the weak topology. 
Suppose that for Smoluchowski's equation with kernel $K=2$, $x+y$ or
$xy$, $\nu$ and $\bar\nu$ are two solutions defined on $[t_0,\Tmax)$,
and make the assumptions stated in the theorem.
Suppose that \qref{R.dto0} fails, i.e., that
\begin{equation}\la{d.notdto0}
\dist( F_t(b(t)\,dx),\bar F_{\bar t}(\bar b(t)\,dx)) \not\to 0 \qquad
\mbox{as $t\to\Tmax$.}
\end{equation} 
Then since $\bar\PP$ is compact, by passing to subsequences we can find
sequences $T_n\uparrow\Tmax$ and $\beta_n=b(T_n)$ and {\em different} 
probability measures $\hat F$, $\check F\in \bar\PP$, such that as
$n\to\infty$ 
we have
\begin{equation}\la{d.seq1}
F_{T_n}(\beta_n x)\to \hat F(x), \qquad
\bar F_{\bar T_n}(\bar \beta_n x)\to \check F(x),
\end{equation}
at every point of continuity of the limit. Here the values $\bar
T_n$, $\bar\beta_n$ are those that correspond via the map 
$(t,b)\mapsto(\bar t,\bar b)$ stated in the theorem. 
Relabeling if necessary, we may assume $0<\hat F(x)$ for some finite
$x$, i.e., $\hat F$ does not represent pure gel.
Therefore, according to the extended \LK\ representation theorem~\ref{ET.LK},
there exists an \barsmeas\ $\dsm$ that corresponds to $\hat F$.

{\em 2.\/} Let 
\begin{equation}\la{d.alp}
\alpha_n = \begin{cases}
\beta_n & (K=2),\cr
\beta_n\rme^{-T_n} & (K=x+y),\cr
\beta_n|T_n| & (K=xy),
\end{cases}
\quad
\lambda_n = \begin{cases}
T_n & (K=2).\cr
\rme^{T_n} & (K=x+y),\cr
|T_n|\inv & (K=xy).
\end{cases}
\end{equation}
and similarly define $\bar\alpha_n$, $\bar\lambda_n$ in terms of
$\bar \beta_n$, $\bar T_n$. Note that $\bar\alpha_n=\alpha_n$.
We claim that $\alpha_n\to\infty$. This is evident for $K=2$, and 
once we prove it for $K=x+y$ it follows for $K=xy$ 
by the transformation formula \qref{M.2}. 
For $K=x+y$, one can prove $\alpha_n\to\infty$ by following the
beginning of the proof of Theorem 7.1 in \cite{MP1} up to (7.8)
using only subsequential convergence. From (7.8) one deduces
$\lambda\rme^{-t}\to \infty$, which corresponds here to
$\alpha_n\to\infty$. 
%[[Ok - desirable to give clear, self-contained proof.]]

{\em 3.\/}  Define rescaled initial \smeass\ (see \qref{R.initG}) by
\begin{equation}\la{d.scG}
\sm\nsup (x) = \lambda_n \alpha_n\inv \sm(\alpha_n x),
\qquad
\bar\sm\nsup (x) = \bar\lambda_n \alpha_n\inv \sm(\alpha_n x),
\end{equation}
According to the extended encoding theorem~\ref{ET.code}
%\marginpar{[[needs statement and proof!]]; added 8/16/06}
the \smeass\ $\sm\nsup$ converge to $\dsm$. 
Let $\vp\nsup$, $\bar\vp\nsup$ and $\Phi$ be the 
first-order Laplace exponents associated
to $\sm\nsup$, $\bar\sm\nsup$ and $\dsm$
respectively as in \qref{L.1}.
We have $\bar\lambda_n = \lambda_n/L(\alpha_n)$, and
with $\hat L(1/q)=\bar\vp(q)/\vp(q)$, the hypothesis
\qref{R.shad2} ensures $\hat L$ is slowly varying and
$\hat L\sim L$.
Hence, for any $q\in(0,\infty)$ we have
\[
\bar\vp\nsup(q) = \bar\lambda_n\bar\vp(q/\alpha_n)
= \vp\nsup(q) 
\frac{\hat L(\alpha_n /q)}{\hat L(\alpha_n)} 
\frac{\hat L(\alpha_n)}{L(\alpha_n)}
\to \Phi(q)
\]
as $n\to\infty$. By Theorem~\ref{LT.Mconv}, it follows
$\bar\sm\nsup$ converges to $\dsm$, and 
by the extended \LK\ representation theorem, 
this yields $\hat F=\check F$ in \qref{d.seq1}. 
This contradicts \qref{d.notdto0} and finishes the proof.
\end{proof}

\subsection{Self similar solutions and domains of attraction}
\label{sec:domains}
% %Here we characterize the domains of attraction 
% [[The idea is to reprove the characterization theorem as an
% application of shadowing, showing how the condition is necessary in
% this case.  ]]
The self-similar solutions are the simplest examples of eternal solutions. 
All self-similar solutions are generated by the \smeass\/ 
$\dsm(x)=C x^{1-\rho}$ with $\rho\in(0,1]$, $C>0$, 
with corresponding Laplace exponents 
\be \la{Do.Hlap}
\Phi(q) = 
Cq^\rho 
\frac{ \Gamma(2-\rho)}{\rho},
\qquad
\Psi(q) = C q^{1+\rho}
\frac{\Gamma(2-\rho)}{\rho(1+\rho)}. 
\ee
Thus, there is a one-parameter family
up to trivial scalings. By \qref{C.eternal} and \qref{A.psi11},
for appropriate $C$, the Laplace exponent $\vp=\vp(t,q)$ of the 
solution satisfies
\be
\vp = \frac{\Phi}{1+t\Phi} = \frac{q^\rho}{1+tq^\rho}
\qquad(K=2),
\ee
\be
q = \vp+\Psi(\rme^t\vp) = \vp + (\rme^t \vp)^{1+\rho}
\qquad (K=x+y).
\ee
The self-similar solutions were described in ~\cite{MP1}, 
and can all be captured by expressing the associated probability distribution 
in the form
\be
F(t,x) = F_{\rho,\gamma}(x/\lgr_{\rho,\gamma}(t)),
\ee
for $\gamma=0,1,2$, where the scale factors are
\be
 \lgr_{\rho,0}(t)=t^{1/\rho},\quad \lgr_{\rho,1}(t)=\rme^{t/\beta}, \quad
\lgr_{\rho,2}(t)=|t|^{-1/\beta}, 
\ee
with $\beta=\rho/(1+\rho)$, and the probability distributions 
$F_{\rho,\gamma}$ are explicitly 
\begin{eqnarray}
\label{eq:f_rho_const}
F_{\rho,0}(x) &=& 
\sum_{k=1}^\infty \frac{ (-1)^{k+1}x^{\rho k}}{\Gamma(1+\rho k)},
\\
\label{eq:f_rho_add}
F_{\rho,1}(x)= F_{\rho,2}(x) &=& 
\frac1\pi \sum_{k=1}^\infty
\frac{(-1)^{k-1}x^{k\beta}}{k!}\Gamma(1+k-k\beta)
\frac{\sin k\pi\beta}{k\pi\beta}.\quad
\end{eqnarray}

We now restate the characterization of the domains of attraction of these
self-similar solutions obtained in~\cite{MP1}. We say a probability measure
on $E$ is nontrivial if it is not concentrated at the origin.

\begin{thm}
\label{DT.domains}
Let $F_t$ denote the probability measure associated to a solution to
Smoluchowski's coagulation equations with $K=2, x+y,$ or
$xy$. 
\bit
\item[(a)] Assume there is a rescaling $b(t)\to \infty$  and a
  nontrivial probability measure $\hat F$   on $E$ 
   such that $F_t(b(t)x)  \to \hat F(x)$ at all points of
   continuity. Then there is $\rho   \in (0,1]$, and a function $L$
   slowly varying at infinity such that 
\be
\la{DE.dom}
\sm_{t_0}(x) =\int_{(0,x]}y^{\gamma+1}\nu_{t_0}(dy) 
\sim x^{1-\rho}L(x), \quad x \to \infty.
\ee 
\item[(b)] Conversely, assume \qref{DE.dom} holds. Then there is a
  rescaling $b(t) \to \infty$ such that 
\be
\la{DE.dom2}
\lim_{t \to \infty} \dist\left(F_t(b(t)\,dx), F_{\rho,\gamma}(dx)
\right)=0. 
%F_{\rho,\gamma}(\lambda_{\rho,\gamma}(t)\,dx)\right) = 0.
\ee
\eit
\end{thm}
Theorem~\ref{DT.domains} illustrates the {\em rigidity of scaling
limits\/}. If we insist on the existence of a proper limit as $t \to
\infty$ (as opposed to subsequential limits), the only possibility is
that $\hat F(x) = F_{\rho,\gamma} (ax)$ for some $\rho \in (0,1]$ and
$a \in (0,\infty)$.  
(For degenerate limits, see the remark below.)
Theorems~\ref{RT.initial_tails} and \ref{RT.shad}
shed more light on  this result as they clarify the main hypothesis
(\qref{DE.dom} above) and allow us to  avoid the use of
Karamata's Tauberian theorem in the proof.  

\begin{proof}
Let us first prove (a). Suppose  there is a (possibly discontinuous)
rescaling $b(t) \to \infty$ such that $\lim_{t \to \infty} F_t(b(t)x) =
\hat{F}(x)$ at all points of continuity of $\hat{F}$. Then $\hat{F}
\in \attr$, so it is associated to a divergent \smeas\/ $\dsm$. 
% We use Theorem~\ref{RT.initial_tails} with $\beta_n=b(T_n)$ where 
% $T_n=n$ ($K=2,x+y$) and $T_n=n^{-1}$ ($K=xy$). 
% % \begin{equation}\label{R.domains1}
% % T_n = \left\{
% % \begin{array}{ll}
% % n/t_0 &(K=2 \text{ or }xy),\\
% %  e^n &(K=x+y). 
% % \end{array}\right.
% % \end{equation}
% % and $\beta_n=b(T_n)$. 
% the measures
% $\tilde{F}^{(t)}(x) = F_t(b(t)x)$, that is 
% \begin{equation}\label{R.domains1}
% a(t) = \left\{
% \begin{array}{ll}
% t/t_0 &(K=2 \text{ or }xy),\\
%  e^t &(K=x+y). 
% \end{array}\right.
% \end{equation}
Theorem~\ref{RT.initial_tails} (ii) now implies the convergence
of the \smeass\/ $\tilde\sm\tsup \to \dsm$ where
\begin{equation}\label{R.dom2}
\tilde\sm\tsup (x) = \tlam\alpha\inv \sm_{t_0}(\alpha x),
\end{equation}
\be \la{Do.alp}
\alpha(t) = \begin{cases}
b(t) & (K=2),\cr
b(t)\rme^{-t} & (K=x+y),\cr
b(t)|t| & (K=xy),
\end{cases}
\quad
\tlam(t) = \begin{cases}
t & (K=2).\cr
\rme^{t} & (K=x+y),\cr
|t|\inv & (K=xy).
\end{cases}
\ee
As we have seen in the proof of Theorem~\ref{RT.shad}, 
$\alpha(t)$ diverges as $t \to \Tmax$ in
each case. Then by \qref{R.dom2}, the Laplace exponent $\vp_0$
for $\sm_{t_0}$ satisfies
\be \la{Do.vps}
\tlam \vp_0(q/\alpha) \to \Phi(q)
\ee
as $t\to\Tmax$, where $\Phi$ is the Laplace exponent of $\dsm$. 
Taking $t\to\Tmax$ along a sequence $t_n$ such that 
$\tlam(t_{n+1})/\tlam(t_n)\to1$, by
a fundamental rigidity lemma~\cite[VIII.8.3]{Feller}, we infer 
that the only possible limits are power-laws, meaning 
$\Phi(q)=cq^\rho$ for some $\rho\ge0$. Since $H$ is a nontrivial
\smeas, we must have $0<\rho\le1$ and $c>0$. 
Moreover we infer $\vp_0$ is regularly varying at 0, meaning
$\vp_0(q)=q^\rho \hat L(q)$, where 
$\hat L(aq)/\hat L(q)\to1$ as $q\to0$ for every $a>0$.
Note that by \qref{Do.vps},
\be\la{Do.tlam}
\tlam \sim c\alpha^\rho/\hat L(1/\alpha),
\quad 
c_n=\tlam(t_n)\vp_0(1/\alpha(t_n))\to c.
\ee

With $t_n$ as described and $\alpha_n=\alpha(t_n)$, 
we claim $\alpha_{n+1}/\alpha_n\to1$ as $n\to\infty$.
Let $a>1$ and suppose $\alpha_{n+1}/\alpha_n>a$ for infinitely many $n$.
Then since $\vp_0$ is strictly increasing,
along this subsequence we have
\[
\frac{c_{n+1}}{\tlam(t_{n+1})}
\frac{\tlam(t_n)}{c_n}
=
\frac{\vp_0(1/\alpha_{n+1})}{\vp_0(1/\alpha_n)}
\le
\frac{\vp_0(a\inv/\alpha_n)}{\vp_0(1/\alpha_n)}
\to a^{-\rho} <1.
\]
But the left-hand side converges to 1. Hence 
$\limsup \alpha_{n+1}/\alpha_n \le1$. Similarly we deduce
$\liminf \alpha_{n+1}/\alpha_n \ge1$, establishing the claim. 

We may now apply the rigidity lemma~\cite[VIII.8.3]{Feller} to \qref{R.dom2} 
to infer that $G_{t_0}$ is regularly varying at $\infty$, meaning
\qref{DE.dom} holds.  (The value of $\rho$ must be the same here, due to
\qref{Do.tlam} and \qref{Do.Hlap}.) This proves part (a).

To prove the converse, we assume that \qref{DE.dom}
holds. Since \qref{DE.dom} holds we may choose 
increasing rescalings  $\alpha(t) \to \infty$ and
$\tlam(t)$ such that the \smeass\/
$G\tsup(x)=\tlam\alpha^{-1} G_{t_0}(\alpha x)$ converge to $\dsm
=x^{1-\rho}$. Let $b(t)$ be defined by \qref{Do.alp} for the
various kernels. It then follows that $F_t(b(t)x) \to \Fgr(x)$ for every
$x >0$. Since the metric is equivalent to weak convergence we also
have \qref{DE.dom2}. 
\end{proof}

\begin{rem} 
A remaining nontrivial possibility discussed in \cite{MP1}
is that of a defective limit on $E$, which we may now take to mean
that $F_t(b(t)x)\to\hat F(x)$ where $\hat F$ is a probability measure 
on $\Ebar=[0,\infty]$, with $0<\hat F(\infty^-)<1$, meaning that gel
appears in the limit.
If this is the case, then $\hat F$ is an element of the full scaling
attractor $\attrx$, and by Theorem~\ref{ET.code}, the rescaled
\smeass\/ $\tilde\sm\tsup\to H$, the \barsmeas\/ associated to
$\hat F$.  Moreover,
$y\inv H(dy)$ must have nonzero charge $h_\infty$ at $\infty$,
and hence $\Phi(0+)>0$. This means that the in the proof above,
the rigidity lemma must yield $\rho=0$, i.e., we must have 
$\Phi(q)=c>0$, corresponding to an eternal extended solution
consisting of a pure dust/gel mixture.
By the discussion in \cite{MP1} (see Remarks 5.4 and 7.4) a necessary and
sufficient condition for this to occur is that
\be 
\int_{[x,\infty)} y\inv \sm_{t_0}(dy) \sim L(x), \quad x\to\infty,
\ee
where $L$ is slowly varying at $\infty$.
\end{rem}

\section*{Acknowledgements}
This material is based upon work supported by the National Science
Foundation under grant nos.\ DMS 00-72609, DMS 03-05985, DMS 06-04420
and DMS 06-05006.  GM thanks
the IMA for partial support during the preparation of the manuscript. 
RLP thanks the DFG for partial support through a Mercator
professorship at Humboldt University.

%\vspace{-6pt}
\pagebreak

\bibliography{mp3}

\end{document}